\documentclass[twocolumn]{revtex4-1}

\usepackage{custompreamble}
\usepackage{tikz}
\usepackage{pgfplots}
\begin{document}
  \title{A Rigorous Demonstration of Superconductivity in a Repulsive Hubbard Model}

  \author{Manjinder Kainth}
  \author{M.W. Long}
  \affiliation{
  School of Physics and Astronomy, University of Birmingham, Edgbaston, Birmingham, B15 2TT,
  United Kingdom.
  }
  \date{\today}
  \begin{abstract}
    We have rigorously shown that a strong Hubbard repulsion can cause superconductivity. The model, which has a particular set of local symmetries, manifests the phase diagram of many unconventional superconductors; anti-ferromagnetism paramagnetism, superconductivity, and even ferromagnetism are all observed. The key technique in the analysis is an exact non-linear fermion transformation, which preserves the correlated motion of electrons while removing the strong interactions. Using resolvent formalism, it is exactly shown that two holes at the Mott point bind to form a localised Cooper pair. As interactions are now weak, we then use BCS mean field theory to calculate the energy, excess pairing, and superconducting gap. These results are compared to exact diagonalisation of finite sized systems and show good agreement. At the Mott point the system is an anti-ferromagnet, and a superconducting phase quickly appears upon doping, and  then vanishes. 

  \end{abstract}
  \maketitle

  \section{Introduction}\label{sec:introduction}
    Superconductivity is the macroscopic quantum phenomena corresponding to zero electrical resistance and perfect diamagnetism, also known as the Meissner effect~\cite{Meissner}.
    The first superconductors were phonon mediated~\cite{Herzfeld1950, Reynolds1950}, described by the formation of Cooper pairs~\cite{Cooper} and subsequent condensation in the BCS solution~\cite{BCS}.
    In these materials Coulomb repulsion can be sidestepped, as the correlation length in these superconductors was large, allowing the electrons to be attracted at a distance where Coulomb is screened.
    Only electrons very close to the Fermi surface participate, within a DeBye energy or so.
    The next generation of superconductors, are strongly correlated and do not fit naturally into this picture. The Coulomb repulsion is dominant and so the interactions are both repulsive and not weak, the coherence length is quite small and so the electrons are not naturally well separated, and there is evidence that all the charge carriers participate and the phenomenon is not restricted to a tiny energy region around the Fermi surface~\cite{IncreasedCarrier}.
    We investigate a strongly correlated model here which is susceptible to superconductivity with these strongly-correlated hallmarks.
    
    The physics of the cuprates is that of doping a Mott insulator~\cite{Lee2006}.
    The Coulomb interaction dominates the chemical bonding and the electrons are localised even though the non-interacting picture would offer a metal. 
    This Mott insulator is usually an anti-ferromagnet, caused by the residual effects of the chemical bonding and the fact that the Coulomb interaction is not infinite; kinetic exchange~\cite{KineticExchange}.
    We will take the mathematical limit that the Coulomb interaction is divergent, eliminating this magnetism, and only reintroduce it as an afterthought; our target is really the superconductivity. 
    Strongly correlated systems are not all superconductors and a variety of phenomena are observed. 
    We also observe a phase which corresponds to the ferromagnetism in the manganites, and the overall picture is a direct competition between this ferromagnet and the superconductor, with the ferromagnet winning in the limit of extreme Coulomb interaction, physically reminiscent of Nagaoka ferromagnetism~\cite{Nagaoka}.
    
    Superconductivity is a tricky property to investigate mathematically. 
    The fundamental issue is that of correlations. 
    In metal physics we know how to describe a non-interacting state, in terms of a Fermi surface and occupancy, but in the presence of interactions we might expect a Fermi-liquid description but we only have vague renormalisation arguments to suggest which correlations might be relevant at low energy. 
    On a more practical level we have mean-field theory, which targets the best non-interacting state to approximate the Fermi-liquid. 
    The positive characteristic of mean-field theory is that it only provides order if the system is susceptible and the negative characteristic is that if the system is susceptible to order then the technique will offer the order even when there are better correlated states available to the system. 
    We will employ the assumption that if the system is susceptible to superconductivity then mean-field theory will predict this and that only low dimensional fluctuations would be expected to destabilise this order via the Mermin-Wagner theorem~\cite{MerminWagner}.
    
    Obviously, mean-field theory is only credible when the interactions are weak, but we are studying a model with divergent repulsion, so we need some non-trivial mathematics to deal with this. 
    Our first step is to map our original strongly correlated Hamiltonian onto another weakly interacting Hamiltonian. 
    This step is exact and is accomplished by a non-linear fermion transformation~\cite{nonlinearfermionMartin}. 
    The resulting description usually has weak interactions and so the mean-field theory should be credible; in addition we have a comparison with an exact diagonalisation study which shows good agreement. 
    Since we are restricted to mean-field theory to demonstrate pairing, we have elected to work in one-dimension. 
    This has the advantage that we can compare with our exact diagonalisation, which is restricted to small systems, but has the disadvantage that long-range phase fluctuations would physically be expected to eliminate any long-range order~\cite{MerminWagner}. 
    The mean-field theory erroneously promotes the long-range order, as it would correctly do in three dimensions, but these weak power-law promoting fluctuations are an irrelevance to the physical interactions which promote the superconducting correlations.
    
    Physically, the mechanism that induces the superconductivity is surprisingly simple. 
    The strong repulsion means that situations with extra local charge have restricted motion, they have to avoid paying the repulsive energy penalty. 
    Situations with less local charge can move around more freely. 
    It can be advantageous to allow local charge fluctuations because the rarified configurations together with the denser blocked regions gain more than the homogeneous average.
    This is depicted in figure~\ref{fig:correlatedmotion}.
    Obviously this requires non-linearity, with the almost vacant being strongly preferred over both the average and the dense configurations. 
    This non-linearity is provided by correlated hopping, chemical bonding that depends on the local occupancy of the site bonded to. 
    This correlated hopping is a generic consequence of non-linear fermion transformations.
    
    Mathematically we employ three independent techniques; non-linear fermion transformations~\cite{nonlinearfermionMartin}, exact diagonalisation~\cite{Lanczos} and resolvent formalism~\cite{ImpurityTheory}. 
    The first technique is the crucial advantage that provides mathematical control. 
    Exact diagonalisation is a standard numerical technique that provides the exact solution to a small finite system. 
    The infinite system is then analysed through finite-size scaling, a form of extrapolation. 
    Resolvent formalism is a technique for finding the exact solution to an eigenvalue problem where, in some basis, there is a trivially completely solvable problem that is only different from the desired problem in its action on a finite number of basis states. 
    For metallic systems, translational invariance controls one particle and then the interactions with a second are local in real space and may be solved using resolvent formalism. 
    We can exactly solve the two-hole problem, allowing a rigorous proof of hole pairing.
    
    The technique of non-linear fermion transformations allows access to a very particular issue. In strongly correlated systems the Coulomb interaction is dominant. 
    Although one electron can naturally occupy a state, a second is strongly repelled by this repulsion and avoids double occupancy. 
    If there is another doubly occupied state with less repulsive losses, then a fermi-liquid can be constructed using quite different single-particle and two-particle states using a non-linear fermion transformation. 
    The choice of such states is usually quite subtle, but we provide an example where there is a unique choice. 
    Note that high temperature superconductivity supplies an excellent example of this problem, one particle occupies a copper orbital but a second sits in an oxygen orbital and forms a Zhang-Rice singlet~\cite{ZhangRice}.
    
    This paper is composed of four parts which, when put together, show that our Hubbard model superconducts.
    In section~\ref{sec:symmetries} with we present the model in question.
    This is a minimal model which encapsulates the physics of unconventional superconductors: a Hubbard model with two atoms per unit cell.
    In the next section we exactly take the physical limit of divergent Coulomb repulsion to constrain the problem to one energy scale and reduce the local state space.
    This is done with a non-linear fermion transformation~\cite{nonlinearfermionMartin} and is the key technique in the analysis.
    Using this, a problem involving divergent energy scales is transformed into one of moderate interactions.
    In the fourth section we provide four key results: the exact binding of two holes in an occupied background, average energy as a function of occupation, excess pair formation, and the superconducting gap.
    Where the results are not exact, we corroborate using exact diagonalisation and will find good agreement.
    All results will then point to this system exhibiting superconductivity.
    In the final section we make physical extensions to the previous work, such as non-diverging Coulomb repulsion, and find that superconductivity is enhanced.

  \section{Model and Local Symmetries}\label{sec:symmetries}
	We tackle a model which is designed to be tractable rather than a model which comes from an experimental system. Our motivation is to gain precise mathematical control and exhibit incontrovertible fact, because the physical idea that we propose, ‘repulsion can lead to superconductivity’, is very controversial and contradicts standard dogma.
	
  	Unconventional superconductors are dominated by two short range interactions: chemical bonding and Coulomb repulsion.
  	The Hubbard model is therefore the natural starting point, due to its elementary treatment of these two interactions.
    We present a minimal Hubbard model that allows for RVB superconductivity.
    It contains two atoms per unit cell permitting singlet formation, and possesses a collection of local symmetries. These will prove crucial in tackling the problem.
    
    The Hamiltonian with which we begin is 
    \begin{multline}
      H = -t_1\sum_{\langle ij \rangle \sigma} (t^{\dagger}_{i\sigma} + b^{\dagger}_{i\sigma}) (t_{j\sigma} + b_{j\sigma}) - t_0 \sum_{i\sigma} (t^{\dagger}_{i\sigma}b_{i\sigma} + b^{\dagger}_{i\sigma}t_{i\sigma}) \\
        + U \sum_{i} (t^{\dagger}_{i\uparrow}t_{i\uparrow} t^{\dagger}_{i\downarrow}t_{i\downarrow} + b^{\dagger}_{i\uparrow}b_{i\uparrow} b^{\dagger}_{i\downarrow}b_{i\downarrow}),
    \end{multline}    
    where \(\langle ij \rangle \) describe neighbouring sites, \( t^{\dagger}_{i\sigma}, t_{i\sigma}, b^{\dagger}_{i\sigma}, b_{i\sigma}\) are independent fermionic creation and annihilation operators with standard anti-commutation relations, \( t_0 \) and \( t_1 \) are hopping parameters and \( U \) is the on site Hubbard repulsion. 
    
\begin{figure}
      \centering      
      \includegraphics[width = \linewidth]{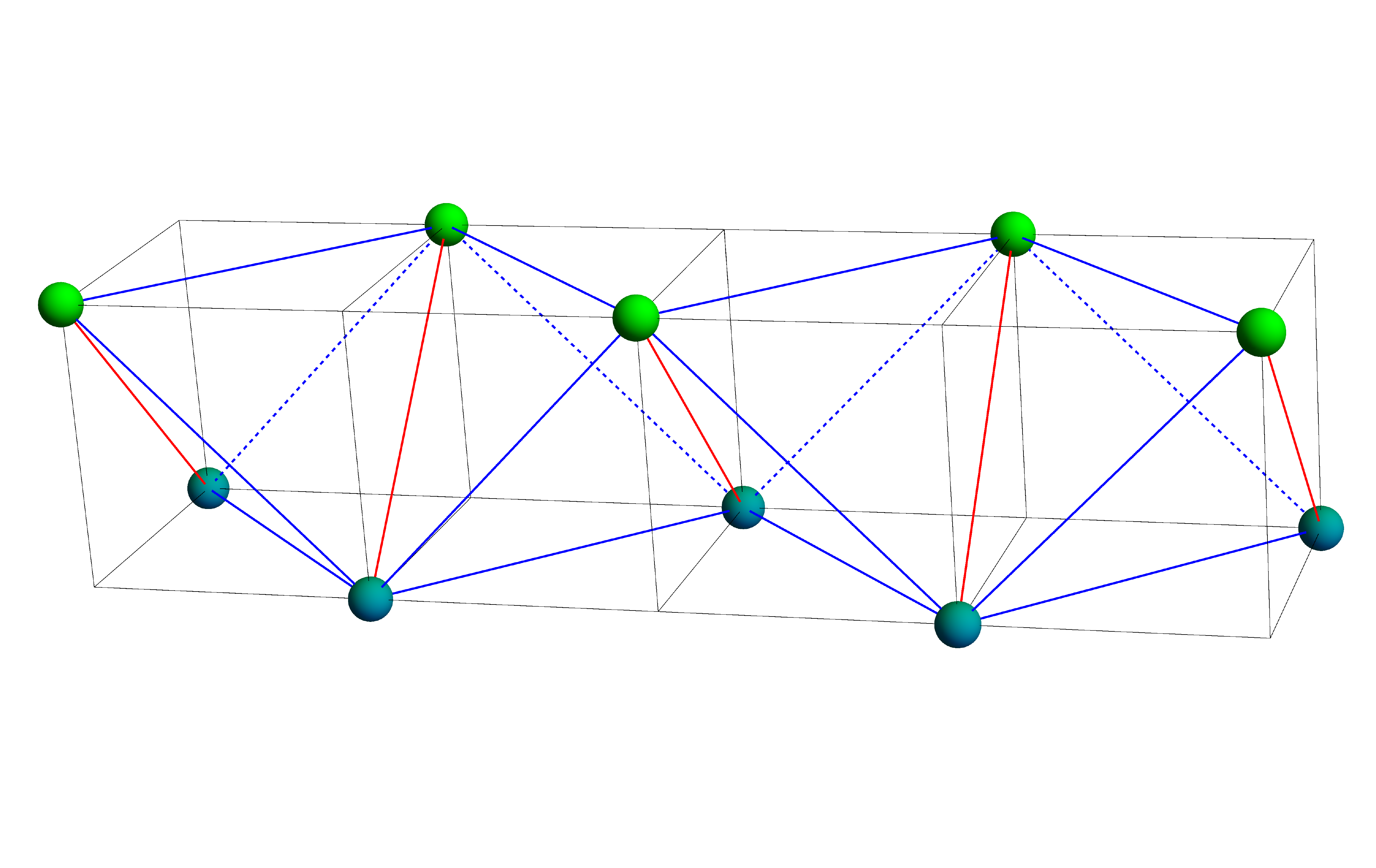}        
      \caption{
        Hamiltonian realisation where operators act in real space, where the lattice is composed of edge sharing tetrahedra (via the red bond).
        Here the green and blue spheres represent the top and bottom sites, with corresponding electron annihilation operators \(t_{i\sigma}\) and \(b_{i\sigma}\). 
        Red and blue lines identify \(t_0\) and \(t_1\) hopping respectively.}\label{fig:hopping1}
    \end{figure}

\begin{figure}
      \includegraphics[width = \linewidth]{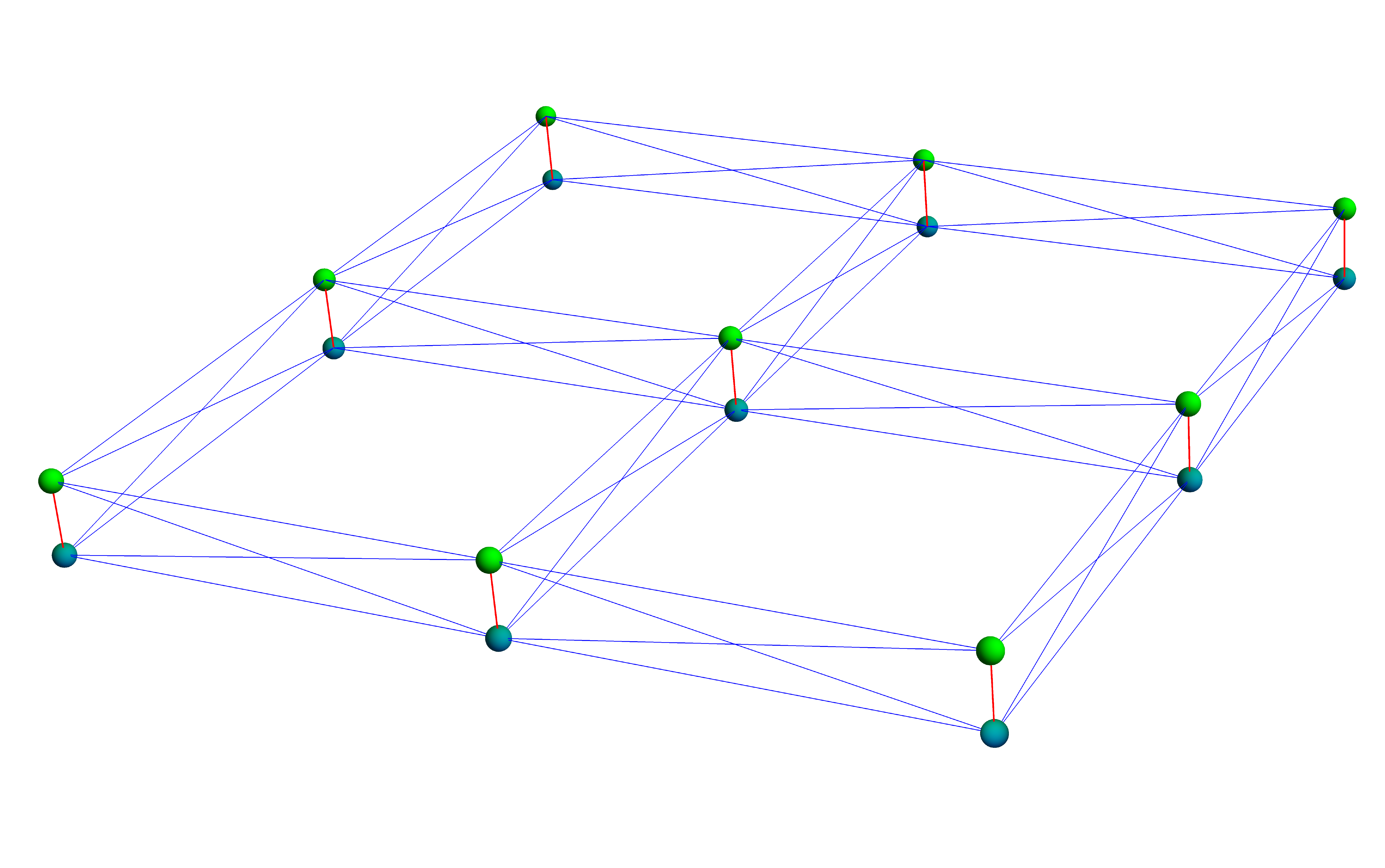}
      \caption{
        Coupled atom realisation of the Hamiltonian where \(t_{i\sigma}\) and \(b_{i\sigma}\) are electron annihilation operators for different orbitals, choice of geometry is arbitrary and lattice index \(i\) corresponds to an individual atom.
        In this case the square lattice is depicted where there is coupling between different orbitals on different sites via the blue \(t_1\) bond, and same sites via the red \(t_0\) bond.
      }\label{fig:hopping2}    
    \end{figure}

    This is a very general model of which there are many physical realisations; we present two such realisations.
    The operators may be attributed to individual sites in real space, and these represent a system of edge sharing tetrahedra via the \(t_0\) bond. This is depicted in figure~\ref{fig:hopping1}.
    Alternatively, the Hamiltonian describes coupling between different orbitals on neighbouring atoms.
    In this case the geometry of the system can be chosen arbitrarily.
    Figure~\ref{fig:hopping2} depicts an example of the latter case where one has modelled a coupled 2D square lattice, mimicking the cuprate layers within YBCO. 
    
  	
\begin{figure}
    	\centering
    	\captionsetup{justification=centerlast}
    	\includegraphics[width = \linewidth]{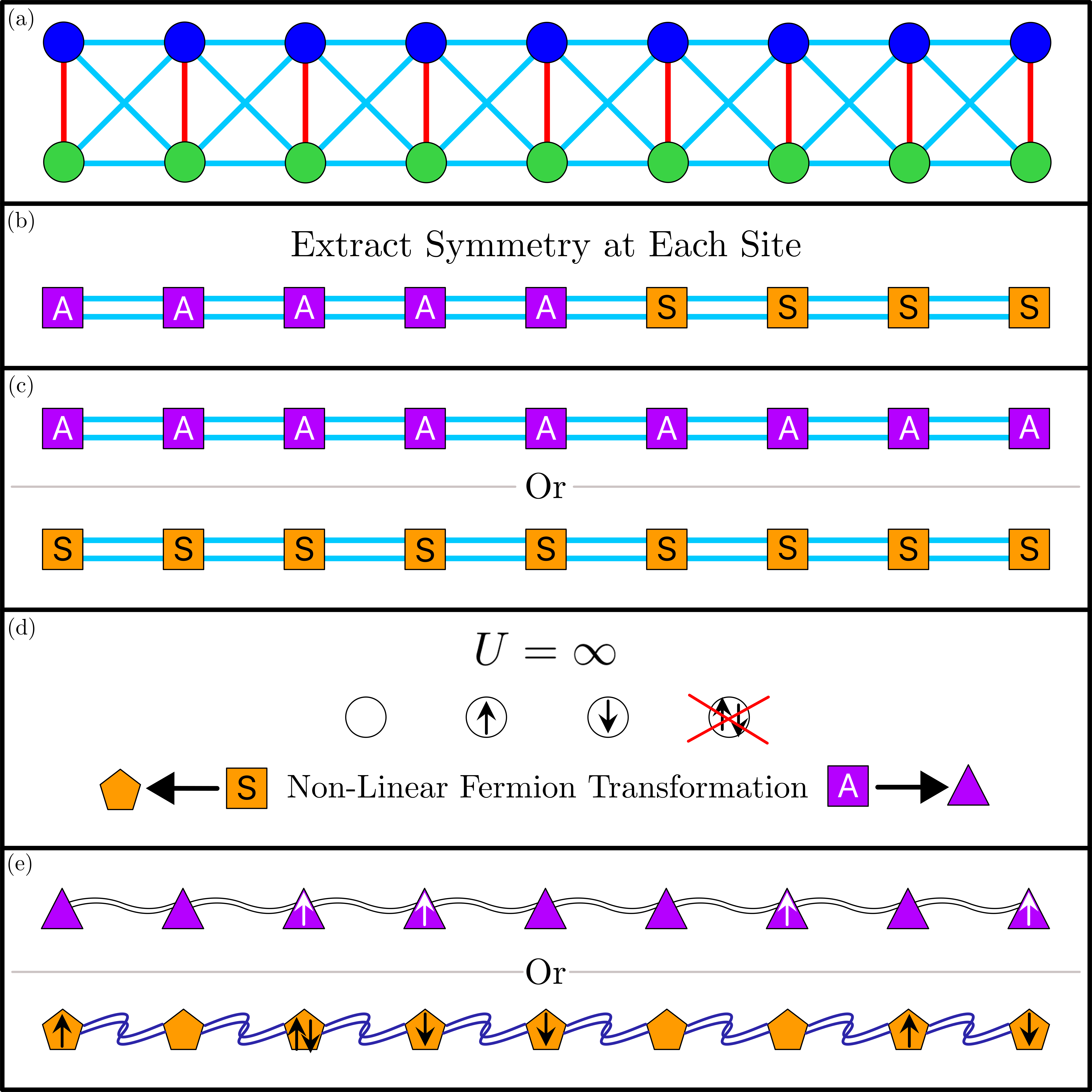}
    	\caption{The process begins with diagram (a), where on any red bond we can switch green and blue sites as the system is symmetric under that transformation. Upon extracting the symmetry we are left with diagram (b), where each site is labelled either `S' or `A' for the symmetry of the state that occupies it. From exact diagonalization results we discover systems are formed of purely `S' or `A', or a phase separated mixture of the two. We arrive at (c) where we examine pure configurations and Maxwell construct where required. In (d) we take the limit \(U = \infty\) which prohibits double occupation in the original basis. This is followed with a non-linear fermion transformation leaving us at (e) with two systems to examine. The original Hamiltonian had 16 degrees of freedom, while the final models have 5 and 4.}\label{fig:symmetry-process}
    \end{figure}

  	We use a local symmetry of this Hamiltonian to drastically simplify the problem, as depicted in figure~\ref{fig:symmetry-process}.
    This symmetry can be thought of as invariance under the local transformation \(t_{i\sigma} \leftrightarrow b_{i\sigma} \).
    Visually this corresponds to an individual green and blue site swapping via the red bond in figures \ref{fig:hopping1} and \ref{fig:hopping2}.
    As diagonal and horizontal bonds are equal in strength this leaves the system unaffected.
    This can be done at each red bond.
    We can label each pair of sites connected by a red bond with `S' (symmetric) or `A' (anti-symmetric) for the symmetry of the local state that occupies them.
    We may then reformulate the problem using operators which are symmetric or anti-symmetric under the aforementioned transformation.
    Explicitly these are
    \begin{equation}
      s_\sigma = \frac{1}{\sqrt{2}} (t_\sigma + b_\sigma),  \quad  a_\sigma = \frac{1}{\sqrt{2}} (t_\sigma - b_\sigma).
    \end{equation}      
    The Hamiltonian can thus be recast as   
    \begin{multline}
      H = -2t_1 \sum_{\langle ij \rangle \sigma} s^{\dagger}_{i\sigma}s_{j\sigma} - t_0 \sum_{i\sigma} (s^{\dagger}_{i\sigma}s_{i\sigma} - a^{\dagger}_{i\sigma}a_{i\sigma}) \\
      + \frac{U}{2} \sum_{i} \Big[ (s^{\dagger}_{i\uparrow}s^{\dagger}_{i\downarrow} + a^{\dagger}_{i\uparrow}a^{\dagger}_{i\downarrow}) (s_{i\downarrow}s_{i\uparrow} + a_{i\downarrow}a_{i\uparrow}) \\
      + (s^{\dagger}_{i\uparrow}a^{\dagger}_{i\downarrow} - s^{\dagger}_{i\downarrow}a^{\dagger}_{i\uparrow}) (a_{i\downarrow}s_{i\uparrow} - a_{i\uparrow}s_{i\downarrow})\Big].
      \label{eq:OGham}
    \end{multline}
    Later we will see that systems are mostly composed of `S' on every site, or `A' on every site.
	Whenever this is not true, a phase separated mixture occurs. 
	This is a regime where the system is split in real space, into two pure regions, with one containing `S' states and the other `A' states.
	This is accurately described by a Maxwell construction~\cite{MaxwellConstruction}, which is discussed in more detail in section \ref{sec:results}, seen in figure~\ref{fig:maxwell-construction}.
    In this paper the systems examined are composed of purely `S' or `A' on each site.
    Henceforth, these are referred to as the symmetric and anti-symmetric subspace.
    
    Utilizing this symmetry greatly simplifies the problem.
    Instead of considering one system with \(16^N\) states we consider two systems with \(8^N\) states, giving us access to larger systems for exact diagonalization.    

  \section{Non-linear fermion transformation}\label{sec:nonlinear} 
    Despite its simplicity, the Hubbard model is notoriously difficult to tackle. 
    The origin of this difficulty is the correlated motion of electrons, which is further complicated by the two competing energy scales.
    However, we can take certain limits, motived by the physics of unconventional superconductors.
    In our case we will take the limit \(U \rightarrow \infty\) motivated by the large on site repulsion to chemical bonding ratio.
    This has two benefits: removing one energy scale, and reducing the local state space in both subspaces. 
    The problem now scales as \(4^N\) and \(5^N\) for the symmetric and anti-symmetric subspaces as opposed to \(8^N\).
    
    If one were to continue using the original basis in this limit, there would be difficulties in remaining within the restricted state space.
    For example, in its current form the Hamiltonian permits the creation of three and four particle states, but the physical limit of \(U \rightarrow \infty\) disallows this.
    To solve this issue we perform a non-linear fermion transformation.
    This is an exact, projection based technique (similar to a Gutzwiller projection) which allows one to transform into any general basis.
    The result is a Hamiltonian which perfectly encapsulates the physics of the problem.
    
    \subsection{Symmetric Subspace}
      In this subsection we will perform a non-linear fermion transformation on the restricted symmetric subspace.
      This will produce a Hamiltonian which acts solely on the four local states permitted by the limit \(U \rightarrow \infty\).
      The operators which arise from this transformation obey standard fermion commutation relations~\cite{nonlinearfermionMartin}, and so are themselves fermions.
      
      The process begins with classifying the local states into two sets, allowed and prohibited.
      The eight local states in the original basis are
      \begin{subequations}
      \begin{align}
        \ket{0}, \quad 
        s^{\dagger}_{\uparrow} \ket{0}, \quad 
        s^{\dagger}_{\downarrow} \ket{0}, \quad 
        \frac{1}{\sqrt{2}}(s^{\dagger}_{\uparrow} s^{\dagger}_{\downarrow} - a^{\dagger}_{\uparrow} a^{\dagger}_{\downarrow}) \ket{0}, \\ 
        \frac{1}{\sqrt{2}}(s^{\dagger}_{\uparrow} s^{\dagger}_{\downarrow} + a^{\dagger}_{\uparrow} a^{\dagger}_{\downarrow}) \ket{0}^\Delta, \quad
        s^{\dagger}_{\uparrow} a^{\dagger}_{\uparrow} a^{\dagger}_{\downarrow} \ket{0}^\Delta, \\
        s^{\dagger}_{\downarrow} a^{\dagger}_{\uparrow} a^{\dagger}_{\downarrow} \ket{0}^\Delta, \quad 
        s^{\dagger}_{\uparrow} s^{\dagger}_{\downarrow} a^{\dagger}_{\uparrow} a^{\dagger}_{\downarrow} \ket{0}^\Delta, \qquad
      \end{align}
      \end{subequations}      
      where the states that are labelled \(\Delta \) are prohibited by the limit \(U \rightarrow \infty \) and hence are projected to zero.
      
      Next we define our new states from the set of allowed ones.
      In principle this choice is arbitrary, but some will be more useful than others.
      We define our new states as
      \begin{subequations}
      \begin{align}
        \ket{0} \equiv \ket{0}, \quad
        c^{\dagger}_{\uparrow} \ket{0} \equiv s^{\dagger}_{\uparrow} \ket{0}, \quad
        c^{\dagger}_{\downarrow} \ket{0} \equiv s^{\dagger}_{\downarrow} \ket{0}, \\
        c^{\dagger}_{\uparrow} c^{\dagger}_{\downarrow} \ket{0} \equiv \frac{1}{\sqrt{2}} (s^{\dagger}_{\uparrow} s^{\dagger}_{\downarrow} - a^{\dagger}_{\uparrow} a^{\dagger}_{\downarrow} ) \ket{0}, \qquad \label{eq:valence}
      \end{align}
      \end{subequations}  
      where \(c^\dagger_{i\sigma}\) and \(c_{i\sigma}\) are standard fermionic creation and annihilation operators, obeying appropriate commutation relations. 
      Note the non-linear nature of the transformation is immediately apparent as \( s^{\dagger}_{\uparrow}\ket{0} = c^{\dagger}_{\uparrow}\ket{0} \) but \( s^{\dagger}_{\uparrow}c^{\dagger}_{\downarrow}\ket{0} \neq c^{\dagger}_{\uparrow}c^{\dagger}_{\downarrow} \ket{0} \).    
      
      The final step is to transform the Hamiltonian.
      This is done by applying the original basis operators to the definitions of our new states.  
      Considering \(s_{\sigma}\) we find
      \begin{subequations}
      \begin{align}
        s_{\sigma} \ket{0} = 0, \quad
        s_{\sigma} c^{\dagger}_{\sigma}\ket{0} &= \ket{0}, \quad
        s_{\sigma} c^{\dagger}_{\bar{\sigma}}\ket{0} = 0, \\
        s_{\sigma} c^{\dagger}_{\sigma} c^{\dagger}_{\bar{\sigma}}\ket{0} &= \frac{1}{\sqrt{2}} c^{\dagger}_{\bar{\sigma}}\ket{0}.
      \end{align}
      \end{subequations}      
      If the action of \(s_{\sigma}\) on a state produces an object which is prohibited by the limit \(U \rightarrow \infty\), that object is set to zero.
      Repeating the process for \(s^\dagger_{\sigma}\) and by appropriately projecting we find
      \begin{equation}
        s_{\sigma} = (1 - \eta c^{\dagger}_{\bar{\sigma}} c_{\bar{\sigma}}) c_{\sigma}, \qquad s^{\dagger}_{\sigma} = (1 - \eta c^{\dagger}_{\bar{\sigma}} c_{\bar{\sigma}}) c^{\dagger}_{\sigma},
      \end{equation}      
      where \( \eta = 1 - \frac{1}{\sqrt{2}} \), is the degree of the non-linearity in \(s_\sigma\).       
      Note \(\eta \approx 0.293\) is moderate in size with comparison to uncorrelated hopping; the divergent nature of \(U \rightarrow \infty\) has been removed while preserving the effects of correlated motion exactly.
      
      Completing this procedure, we find the symmetric subspace Hamiltonian to be 
      \begin{multline}
        H_S = -2t_1 \sum_{\langle ij \rangle \sigma} (1 - \eta c^{\dagger}_{i\bar{\sigma}} c_{i\bar{\sigma}}) c^{\dagger}_{i\sigma}  c_{j\sigma} (1 - \eta c^{\dagger}_{j\bar{\sigma}} c_{j\bar{\sigma}}) \\
        - t_0 \sum_{i\sigma} c^{\dagger}_{i\sigma} c_{i\sigma} + 2t_0 \sum_i c^{\dagger}_{i\uparrow} c_{i\uparrow} c^{\dagger}_{i\downarrow} c_{i\downarrow}.
        \label{eq:symham}
      \end{multline} 
      At this point we note a few things.
      From equation \eqref{eq:valence}, we see the original Mott point (which is mapped to full occupation in the new basis) is composed of valence bond states.
      This is the only way to avoid the Coulomb penalty for a two particle state.
      Second, the resulting Hamiltonian is similar to a tight binding model but it takes into account correlated motion, the size of which is moderate when compared to uncorrelated motion.
      With the application of physical limits and a non-linear fermion transformation, we have mapped a complex problem with two differing energy scales onto a simplified problem containing the correct physics and one energy scale.
      
\begin{figure}
        \centering
        \includegraphics[width = \linewidth]{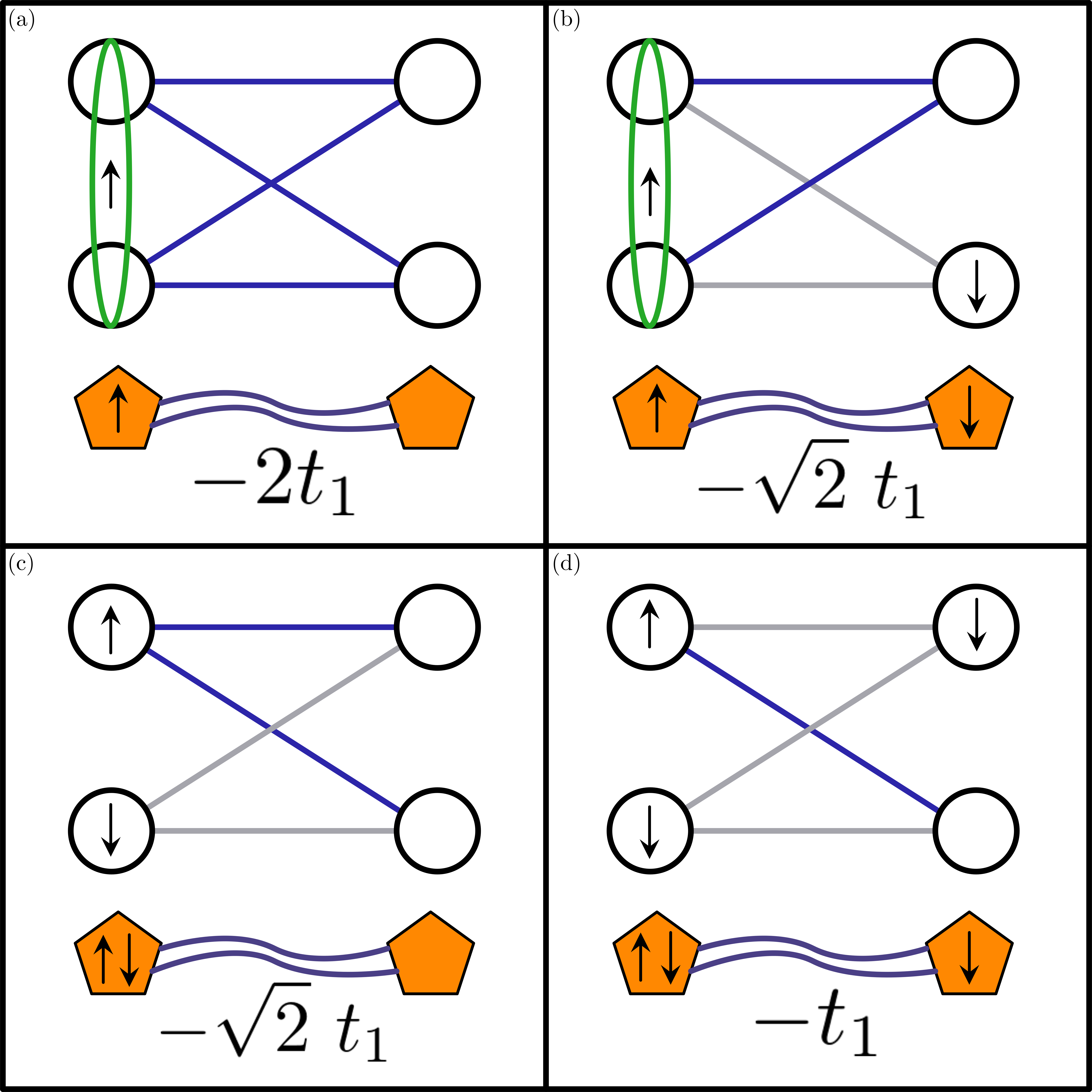}
        \caption{
          The energy gain for hopping a particle with spin \(\uparrow\) from the left site to the right.
          Accessible hops are highlighted in blue.
          Green rings describe a superposition over both top and bottom sites.
          At low occupation the system is dominantly described via (a) and (b), where divergent Coulomb penalises particles next to one another with a decreased hopping.
          However as occupation increases the system must decide between (b) or (c) and (d). 
          With the latter being penalised the system has no choice but to form pairs and keep them closely bound (c and d) to circumvent the Coulomb penalty.
        }\label{fig:correlatedmotion}
      \end{figure}

      In section~\ref{sec:results} it will be shown that superconductivity is mediated by holes close to the Mott point.
      For this reason we now examine the behaviour of holes in the \(c_{i\sigma}\) basis.
      A single hole moves through the occupied background with hopping \(t_1\), whereas two holes (paired or otherwise) does so with \(\sqrt{2}t_1\).
      Moreover three holes move with hopping \(2 t_1\).
      This is depicted in figure \ref{fig:correlatedmotion}.
      In this subspace, holes are being drawn together via an effective attractive interaction: circumvention of the Coulomb penalty.
      It is this effect which drives superconducting pairing.
      The exact mathematical nature of this is discussed in section \ref{sec:analysis}.
      
    \subsection{Anti-Symmetric Subspace}
      In this subsection we attempt to repeat the previous process for the anti-symmetric subspace.
      Unfortunately it is not as effective as in the previous subsection.
      Instead we must extend the process with the use of physical arguments and numerical results.
      
      The eight local states in this subspace are 
      \begin{subequations}
      \begin{align}
        a^{\dagger}_{\uparrow} \ket{0}, \quad
        a^{\dagger}_{\downarrow} \ket{0}, \quad
        s^{\dagger}_{\uparrow} a^{\dagger}_{\uparrow} \ket{0}, \quad 
        s^{\dagger}_{\downarrow} a^{\dagger}_{\downarrow} \ket{0}, \qquad \\ 
        \frac{1}{\sqrt{2}}(s^{\dagger}_{\uparrow} a^{\dagger}_{\downarrow} + s^{\dagger}_{\downarrow} a^{\dagger}_{\uparrow}) \ket{0}, \quad
        \frac{1}{\sqrt{2}}(s^{\dagger}_{\uparrow} a^{\dagger}_{\downarrow} - s^{\dagger}_{\downarrow} a^{\dagger}_{\uparrow}) \ket{0}^\Delta, \\
        s^{\dagger}_{\uparrow} s^{\dagger}_{\downarrow} a^{\dagger}_{\uparrow} \ket{0}^\Delta, \quad
        s^{\dagger}_{\uparrow} s^{\dagger}_{\downarrow} a^{\dagger}_{\downarrow} \ket{0}^\Delta, \qquad \quad
      \end{align}
      \end{subequations}      
      where the prohibited states are labelled with \(\Delta \).    
      
      The anti-symmetric subspace requires an \(a^\dagger_\sigma\) on each site and therefore only exists for occupancies greater than \(N/2\).      
      The usefulness in the non-linear fermion transformation is mapping a problem to ensure local state space restriction.
      In the previous case we were left with four states, and this is conveniently the same number of states for spin-half fermionic system.
      To do this the number of local states must be decomposable, unfortunately in this case the number of remaining states (five) cannot be.      
      Note that this is usually the case, and highlights how fortunate we were in the case of the symmetric subspace.
      In order to create an effective theory we consider the energetics of the system.
      From equation \eqref{eq:OGham} we see an \( a^{\dagger}_{\sigma} \) is localised to a site, while the \( s^{\dagger}_{\sigma} \) particles are free to move.      
      In the anti-symmetric subspace all one particle states are spin 1/2 and two particle states are spin one.
      Consider two neighbouring sites with three particles between them.
      As one site has two particles and the other has one particle, the total spin of the pair must be either 3/2 or 1/2.
      When the \( s^{\dagger}_{\sigma} \) particle hops to its neighbouring site, it does so with \(-2t_1\) if the total spin of the pair is 3/2 and \(-t_1\) if the total spin is 1/2.
      The energetics signify that the anti-symmetric subspace must therefore be an itinerant ferromagnet.
      Upon examination of numerical results, discussed in section \ref{sec:results}, we see that the total spin is maximal in all tested cases.
      This is the physical manifestation of Nagaoka ferromagnetism~\cite{Nagaoka} on bipartite lattices in our system.
      
      Given the ferromagnetic nature of the system, the Hamiltonian for the anti-symmetric subspace is given by
      \begin{equation}\label{eq:AS-ham}
        H_{A} = -2t_1 \sum_{\langle ij \rangle} s^{\dagger}_{i\uparrow} s_{j\uparrow} - t_0 \sum_{i} \left(1- s^{\dagger}_{i\uparrow} s_{i\uparrow} \right),
      \end{equation}  
      where the vacuum is composed of an \(a^\dagger_\uparrow\) on each site.    
      
      In this section we took the physically motivated limit \(U \rightarrow \infty\). 
      This reduced the local state space, allowing us to use a non-linear fermion transformation to advance the problem.
      The result was two Hamiltonians, equations \ref{eq:symham} and \ref{eq:AS-ham}, acting on the two subspaces being investigated, each in a position to be analysed in the following section.

  \section{Analysis}\label{sec:analysis}
    In noteworthy physical problems, the Hamiltonian is rarely solved trivially.
    For many cases exact solutions do not exist and hence numerics or approximate techniques are used.
    In this paper if there is an exact solution we will use it, if not we will perform exact diagonalisation and approximate analysis.
    We find the anti-symmetric subspace is exactly described as an itinerant ferromagnet, while the symmetric subspace has a superconducting phase for a region of the phase diagram.
    We begin with examining one dimensional systems as our exact diagonalization results have better finite-size scaling in 1D.
    Though superconductivity is not permitted in one dimension due to the Mermin-Wagner theorem, quasi-long range order is. Hence, which provides qualitatively similar results may be found.
    Regardless, this work is extended to 2D in section~\ref{sec:physicalExtensions} where this is not an issue and superconductivity persists at zero temperature.
    
    The origin of conventional superconductivity is Cooper pair formation; these pairs then form the basis of the BCS solution.
    Following a similar structure, we first prove that real space hole pairs form, and use these as the basis of our mean field solution.
    
    \subsection{Pair Formation}
      In a free electron gas, the Fermi surface is unstable to pair formation due to interactions with phonons.
      These Cooper pairs form in momentum-space and hence have a large correlation length.
      This theory is insufficient for unconventional superconductors where the energy scale is much larger than the Debye frequency and the correlation length is small.
      We exactly examine a system composed of two holes in an occupied background and find they form a localised pair, where the energy scale is the chemical bonding.
      This is done using resolvent formalism.
      
      The exact solution works as follows. 
      First split the Hamiltonian in two: an exactly solvable component \(H_0\) and an `impurity' \(H_1\) that only affects a small number of states \(H = H_{0} + H_{1}\).      	
      This method relies on understanding and dealing with the resolvent \(G(\epsilon) = (\epsilon - H)^{-1}\). 
      Using completeness this can be rewritten as
      \begin{equation}
	      G(\epsilon) = \sum_{n} \frac{\ket{\psi_n}\bra{\psi_n}}{\epsilon - E_n}.
      \end{equation}
      Note that there are poles at the eigenvalues of \(H\), whose residues are their corresponding eigenfunctions.
      By defining \(G^0(\epsilon) = (\epsilon - H_0)^{-1}\) it can be shown that 
      \begin{equation}\label{eq:expanded-resolvent}
	      G(\epsilon) = G^{0}(\epsilon) + G^{0}(\epsilon) \Sigma(\epsilon) G^{0}(\epsilon),
      \end{equation}
      where \(\Sigma(\epsilon) = H_1 (1 - G^{0}(\epsilon)H_1)^{-1}\).
      From equation~\ref{eq:expanded-resolvent} we can see that the poles, and hence the energy eigenvalues, of \(G(\epsilon)\) are either poles of \(G^0\) or of \(\Sigma(\epsilon)\).
      We are only interested in `new' poles as they correspond to energies due to the addition of \(H_1\).
      Therefore, we must calculate the poles of \(\Sigma(\epsilon)\) which is equivalent to to solving the eigenvalue equation
      \begin{equation}
      	\ket{\Phi} = G^0(\epsilon) H_1 \ket{\Phi}.
      \end{equation} 
      This calculation involves a finite dimensional inverse, controlled by the number of states affected by \(H_1\), and is tractable for small matrices.
      The two independent methods to calculate \(G^0(\epsilon)\), the sum and the eigenvalue equation, allow us to equate the two and solve for \(\epsilon\).
      If this energy is lower than the ground state energy of \(H_0\), there is a bound state.
      
      Our symmetric subspace Hamiltonian \eqref{eq:symham} can be separated in the prescribed manner.
	  The system we examine is composed of two holes with opposite spin, centre-of-mass momentum \(q\), separated by \(m\) sites.
      \begin{equation}
	      \ket{m}_q = \frac{1}{\sqrt{N}}\sum_{j} e^{iq(j + \frac{m}{2})} c_{j,\uparrow} c_{j+m,\downarrow} \ket{\Psi_{\mathrm{Mott}}},
	      \label{eq:holeState}
      \end{equation}
      where the set of states \(\ket{m}_q\) form an orthonormal basis as required.             
      The Hamiltonian can be written as \(H_S = H_0 + H_1\), where 
      \begin{equation} \label{eq:impurity-h0}
        H_0 = -2t_1 \sum_{\langle ij \rangle \sigma} c^{\dagger}_{i\sigma} c_{j\sigma} - t_0 \sum_{i\sigma} c^{\dagger}_{i\sigma} c_{i\sigma},
      \end{equation}
      is simply the tight binding model with an on site interaction, solved by a Bloch transformation.
      The matrix elements of \(H_1\) is only non zero for states \(\ket{1}_q, \ket{0}_q, \ket{\bar{1}}_q\), and is given by 
      \begin{equation}
	      H_1 = 
	      \begin{bmatrix}
	      0& \kappa& 0 \\
	      \kappa& 2t_0& \kappa \\
	      0& \kappa& 0 
	      \end{bmatrix}
	      \label{eq:impurityMatrix}
      \end{equation}
      where \(\kappa = 2\sqrt{2}t_1\eta\cos(\frac{q}{2})\).     
      
\begin{figure}
      	\centering
      	\captionsetup{justification=centerlast}
      	\includegraphics[width = \linewidth]{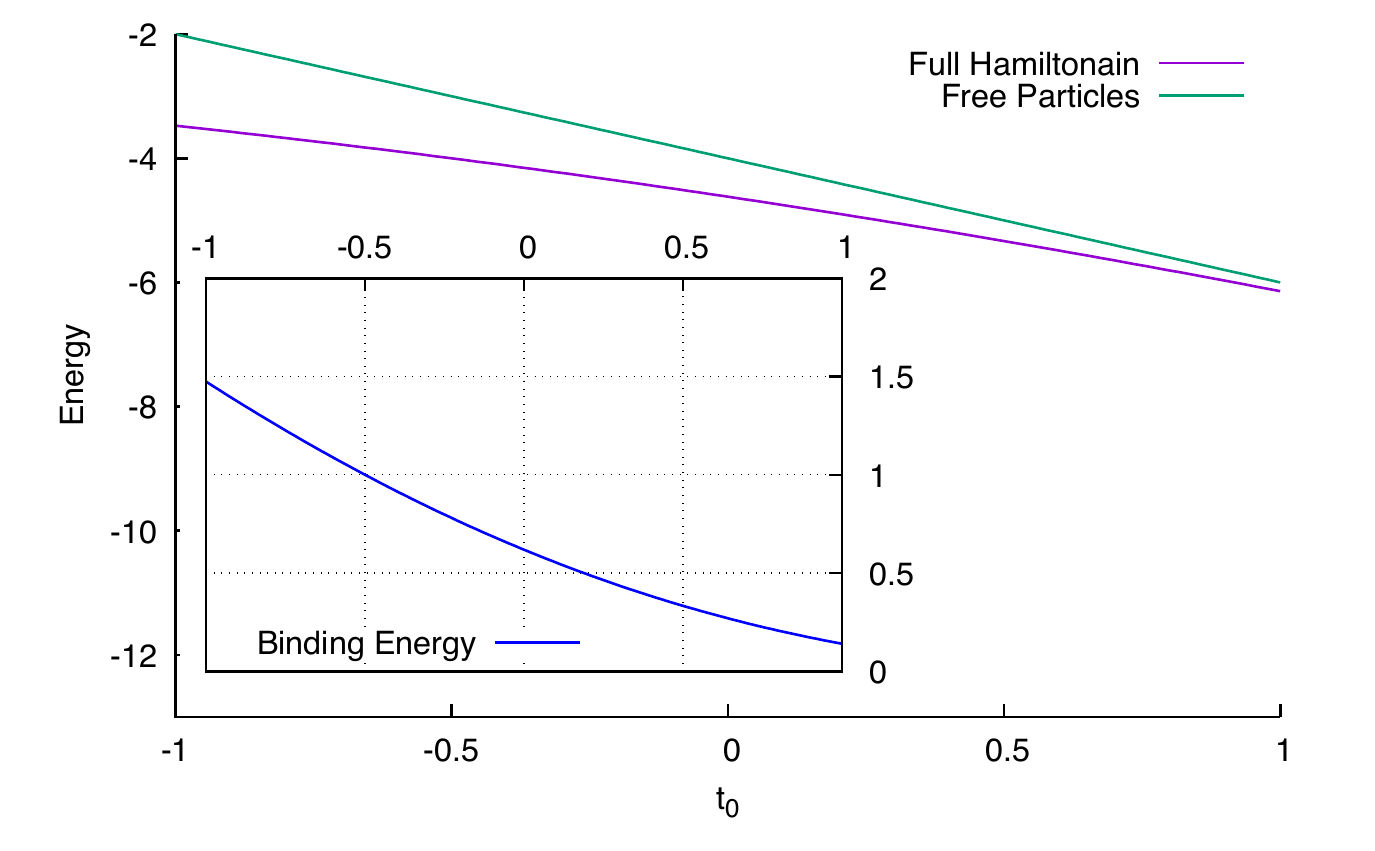}
      	\caption{Total energy for two particles with Hamiltonian~\ref{eq:impurity-h0}, labelled free particles, and exact solution labelled full Hamiltonian. The difference of the two is the binding energy.}\label{fig:convergence-1d}
      \end{figure}

      This calculation can be performed for finite sized systems --- where they match diagonalisation results to numerical accuracy --- or in the continuum limit, where the energy of the pair of holes is given by 
      \begin{equation}
        \epsilon(q) = \frac{4}{3}\left(-t_0 - \sqrt{t_0^2 + 12t_1^2\cos^2\left[\frac{q}{2}\right]}\right).
      \end{equation}      
      This only describes a bound state if the energy is lower than the free particles, giving the constraint
      \begin{equation}
        t_0 > - \frac{t_1\cos(\frac{q}{2})}{2},
      \end{equation}      
      assuming \(t_1 > 0\). This is depicted in figure~\ref{fig:convergence-1d}.
            
\begin{figure}
        \centering
        \captionsetup{justification=centerlast}
        \includegraphics[width = \linewidth]{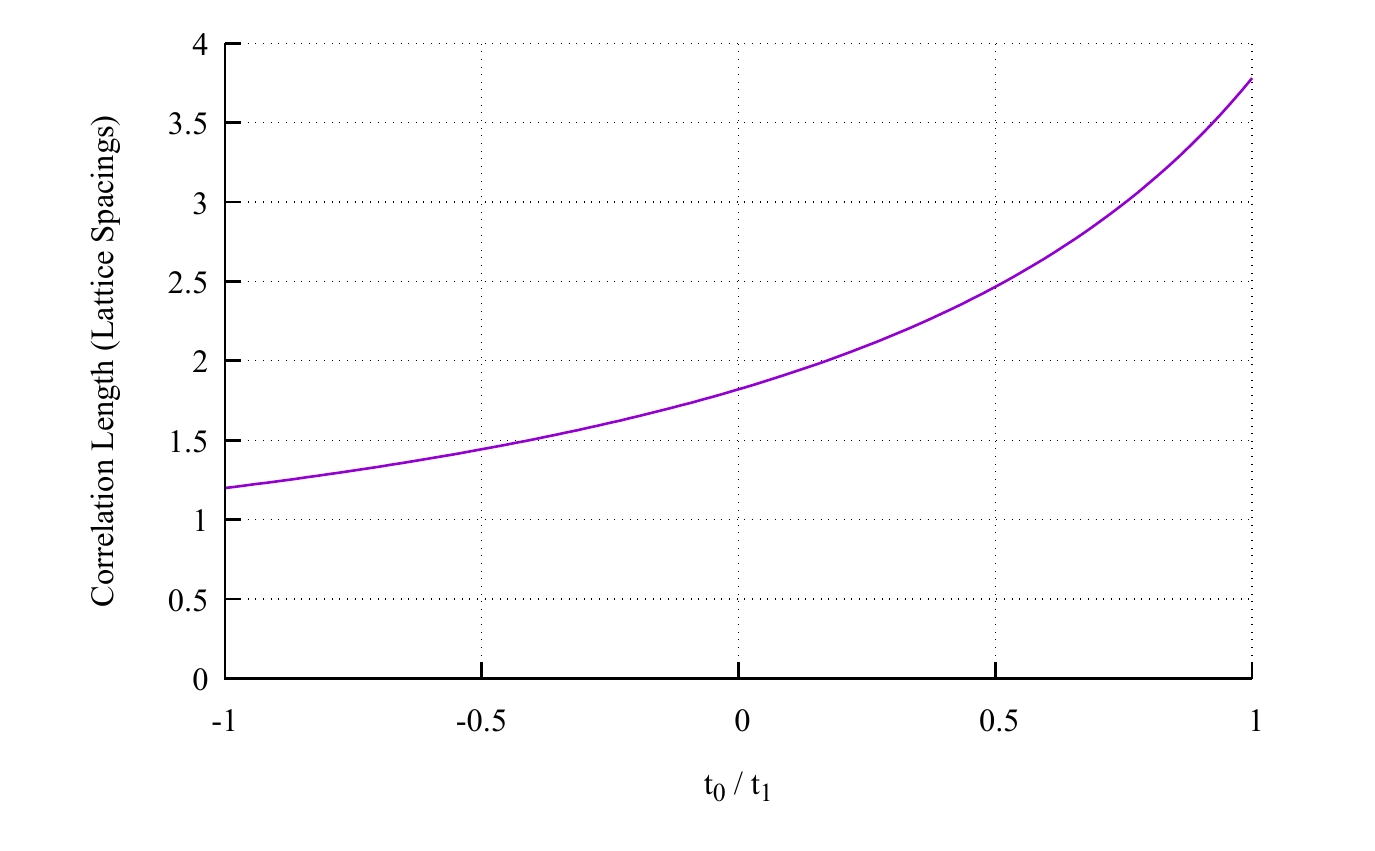}
        \caption{Corelation length of the bound pair of holes as a function of \(t_0/t_1\) measured in lattice spacings.}\label{fig:correlation-length}
      \end{figure}

      This method also gives us the wavefunction of the bound state, and using this we find the correlation length of the pair to be
      \begin{equation}
      	\frac{1}{\xi} = \mathrm{ln}\left[\frac{6}{\tau + \sqrt{\tau^2 + 12}}\right], \quad \tau = \frac{t_0}{t_1}.
      \end{equation}
      This is depicted in figure~\ref{fig:correlation-length}. 
      The pair is very closely bound.
      
      Exact calculations with more holes in the system are not tractable, instead we must rely on BCS mean-field theory with these pairs forming the basis.
      Note in our system the interactions are smaller than the hopping parameter \(t_1\), and so mean-field theory is a valid technique to use.
      To provide credence to the validity of the mean field theory we perform exact diagonalisation calculations on finite sized systems and compare results in section~\ref{sec:results}.
      
    \subsection{Mean Field Theory}
      What follows is a BCS solution of \(H_S\) using Hartree-Fock mean field theory.
      The nature of this solution is the same as BCS, however the work is akin to that of Valatin~\cite{Valatin}; consequently this section can be skipped if desired.
        
      The assumptions in this analysis are translational invariance, paramagnetism, spin invariance and zero temperature.
      The permitted correlations are
      \begin{subequations}
        \begin{align}
          n_0 = \langle c^{\dagger}_{i,\sigma}c_{i,\sigma} \rangle, \qquad n_1 = \langle c^{\dagger}_{i,\sigma}c_{i+1,\sigma} \rangle,\\
          \delta_0 = \langle \sigma c^{\dagger}_{i\sigma} c^{\dagger}_{i\bar{\sigma}} \rangle, \qquad
          \delta_0^{*} = \langle \sigma c_{i\bar{\sigma}} c_{i\sigma} \rangle, \quad \\
          \delta_1 = \langle \sigma c^{\dagger}_{i\sigma} c^{\dagger}_{i+1\bar{\sigma}} \rangle, \quad
          \delta_1^{*} = \langle \sigma c_{i\bar{\sigma}} c_{i+1\sigma} \rangle ,
        \end{align}
      \end{subequations}       
      where \(n_0\) and \(n_1\) are the on-site and nearest-neighbour occupation, \(\delta_0\), \(\delta_0^{*}\),  \(\delta_1\), \(\delta_1^{*}\) are on-site and nearest-neighbour superconducting pair occupation and \(\sigma\) is used as both the spin index \(\sigma = \uparrow \) or \(\downarrow\) and \(+\) or \(-\).
      Without loss of generality we choose a phase such that \(\delta_0 = \delta_0^{*}\) and \(\delta_1 = \delta_1^{*}\).      
      We find the superconducting average energy per spin per site to be 
      \begin{multline}
        \bar{E}_{SC} = -8t_1 n_1 ((1-\eta n_0)^2 - \eta^2 n_1^2) - 2t_0 n_0 (1 - n_0) \\ 
        + 2t_0\delta_0^2 + 8t_1 (2\eta \delta_0 \delta_1 (1-\eta n_0) + \eta^2 n_1 (\delta_0^2 + \delta_1^2)),
      \end{multline}    
      while the paramagnetic average energy is given by \(\bar{E}_{P} = \bar{E}_{SC}|_{\delta = 0}\).
      In order to provide a self-consistent definition of \(n_0, n_1, \delta_0\) and \(\delta_1\) we use Wick's theorem~\cite{Wick} to find an effective single particle Hamiltonian, and diagonalize it using a Bogilubov-Valatin transformation~\cite{Valatin}.
      This results in a gapped dispersion given by 
      \begin{equation}
        E_k^{\pm} = \pm \sqrt{A_k^2 + B_k^2}.
      \end{equation}    
      where \(A_k = \alpha + \beta \gamma_k - \mu \), \(B_k = \nu + \lambda\gamma_k \), and \(\gamma_k = \cos(k)\) is the structure factor for a 1D chain and \(\mu\) is the grand canonical chemical potential controlling the number of particles in the system. Here \(\alpha, \beta, \nu, \lambda\) are given by
      \begin{subequations}
      \begin{align}
        \alpha &= 2t_0(2n_0 - 1) + 16t_1\eta (n_1(1-\eta n_0) - \eta\delta_0 \delta_1), \\
        \beta &= -8t_1 ((1-\eta n_0)^2 - \eta^2(3n_1^2 + \delta_0^2 + \delta_1^2), \\
        \nu &= 4t_0\delta_0 + 16t_1\eta(\delta_1(1-\eta n_0) + \eta n_1 \delta_0), \\
        \lambda &= 16t_1\eta(\delta_0(1-\eta n_0) + \eta n_1 \delta_1).
      \end{align}
      \end{subequations}  
      Note the size of the gap is of the order of the hopping parameter, hence the transition temperature of this system is high --- similar to that of many unconventional superconductors.
      Using the diagonal operators we find the following self consistent integrals 
      \begin{subequations}
      \begin{align}
        n_0 &= \int_{-\pi}^{\pi} \frac{\mathrm{d}k}{2\pi} \frac{1}{2} \left(1 - \frac{A_k}{E_k^{+}}\right),  
        &&\delta_0 = \int_{-\pi}^{\pi} \frac{\mathrm{d}k}{2\pi} \left(\frac{-B_k}{2E_k^{+}}\right), \\
        n_1 &= \int_{-\pi}^{\pi} \frac{\mathrm{d}k}{2\pi} \frac{1}{2} \left(1 - \frac{A_k}{E_k^{+}}\right)\gamma_k,
        &&\delta_1 = \int_{-\pi}^{\pi} \frac{\mathrm{d}k}{2\pi} \left(\frac{-B_k}{2E_k^{+}}\right) \gamma_k.
      \end{align}
      \end{subequations}      
      Varying \(\mu\) and numerically calculating the self-consistent integral provides a numerical value for each of the parameters, which in turn are used to calculate \(\bar{E}_{SC}\).
      
  \section{Results}\label{sec:results}
    In section~\ref{sec:analysis} we showed that two holes in a fully occupied background bind, and a collective group of these form a superconducting solution.
    Now we show the validity of these results and inspect other properties of superconductivity: pair formation and the superconducting gap.
    The validity of our results are demonstrated by comparison with exact diagonalisation of finite sized systems, where there is good agreement between numerical and analytic results.
    
    \subsection{Phase Separation and Maxwell Construction}
    To begin with we must consider the system as a whole.
    In section~\ref{sec:symmetries} we argued that we can consider pure symmetry configurations,.
    Now we shall examine how they interact.    
    Consider a system with \(N\) lattice points, where each point is labelled either `S' or `A' for the symmetry of the state that occupies it.
    For each value of \(t_0\) and occupation \(n_0\) there exists a ground state configuration of S's and A's. 
    Using exact diagonalisation we are able to extract this symmetry and find the configuration. 
    We find there are only two styles of configuration:
    Either purely one symmetry or a phase separated mixture. 
    This is where the configuration is split in two, with one region containing all the A's and the other S's.
    Analytically we can calculate the energy of a phase separated mixture using a Maxwell construction~\cite{MaxwellConstruction}.
    This is depicted in figure~\ref{fig:maxwell-construction}. 
    The exact diagonalisation of the full system perfectly coincides with the pure symmetry systems in the appropriate regions. 
    Where it does not, the agreement with the Maxwell construction is impeccable.
    Therefore, we conclude that we can understand the physics of the system as a whole by combining the results from each subspace.       
    This model does not take into account long range Coulomb forces, but this can be added qualitatively. 
    In our system each phase in the mixture contains different electron numbers.
    Therefore, in a real material, creating a fully phase separated state would incur a massive Coulomb penalty; to mitigate this the separation would occur instead on the micro- or meso-scopic scale.
    This has been seen experimentally in Sr\(_{0.5}\)Ce\(_{0.5}\)FBiS\(_{2-x}\)Se\(_x\) where ferromagnetism and superconductivity were shown to coexist~\cite{macroPhaseSeperation}.
    
    \subsection{Energy}
\begin{figure}
      \centering
      \includegraphics[width = \linewidth]{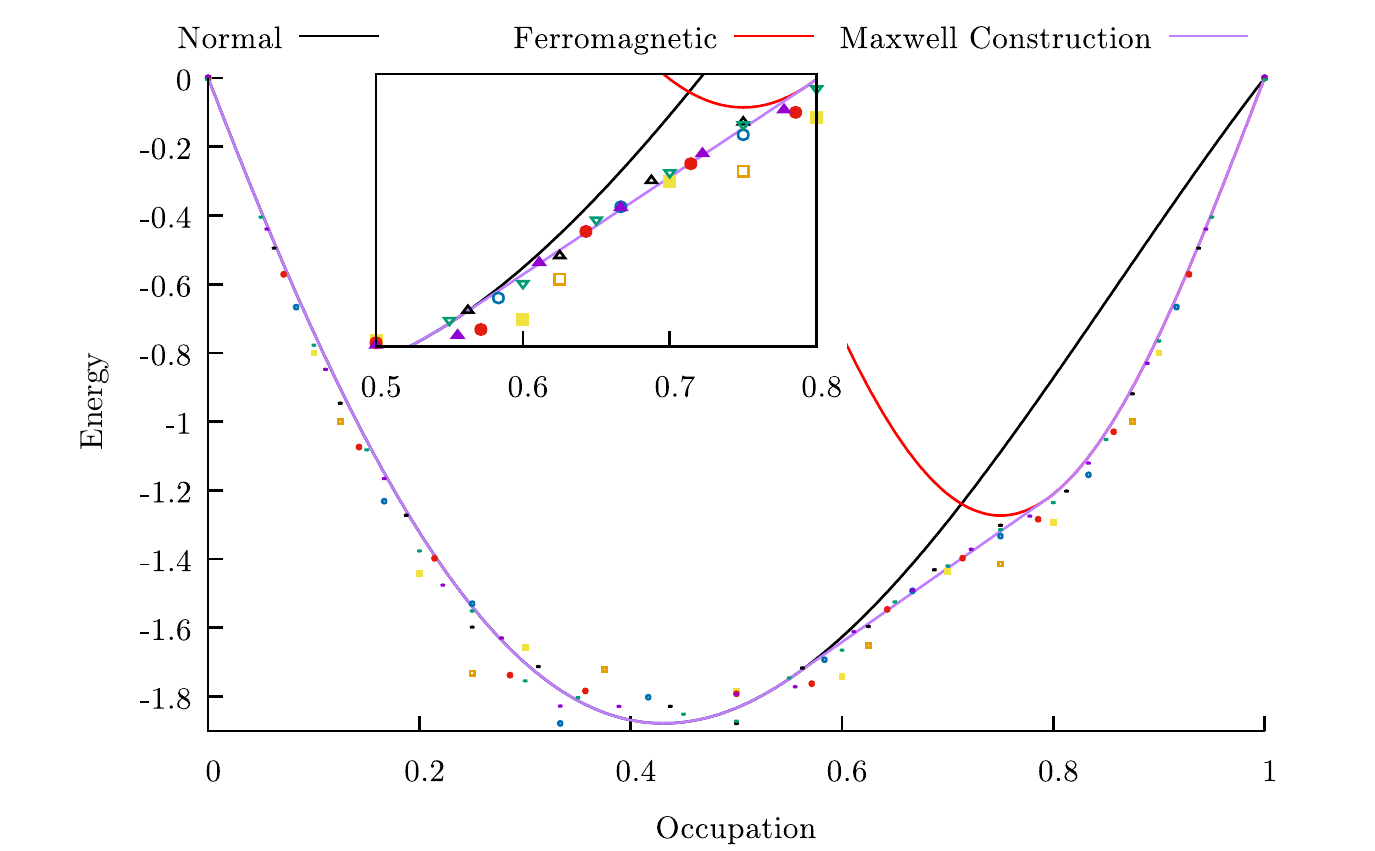}
      \caption{Energy per site per spin as a function of occupation, comparing mean field theory to diagonalisation of finite systems of size 8 to 10 with \(t_0 = 0\) with \(t_1 = 1\).       
      The three systems examined are the full system, symmetric, and anti-symmetric.
      Predominantly the pure symmetry configurations energies match the full system.
      When these systems are in competition a Maxwell construction between the subspaces shows good agreement to the data.}\label{fig:maxwell-construction}
    \end{figure}

\begin{figure}
      \centering
      \includegraphics[width = \linewidth]{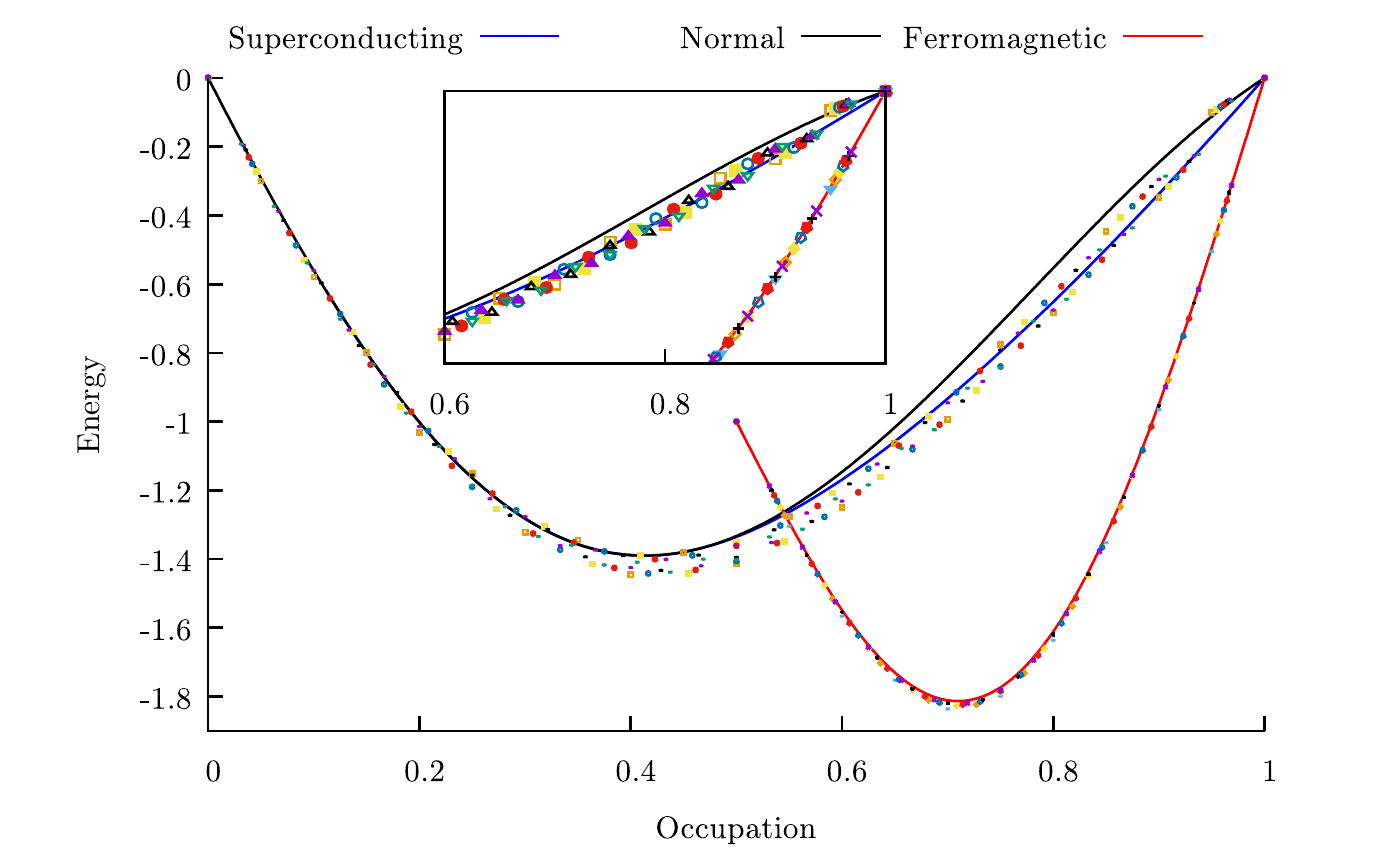}
      \caption{Energy per site per spin as a function of occupation, comparing mean field theory to diagonalisation of finite systems of size 10 to 14 with \(t_0 = -t_1\) with \(t_1 = 1\). 
      Superconducting mean field theory, existing within the symmetric subspace, provides better agreement for diagonalisation results.
      Ferromagnetism is the phase within the anti-symmetric subspace and is energetically dominant over superconductivity.}\label{fig:energy1}
    \end{figure}

\begin{figure}
      \centering
      \includegraphics[width = \linewidth]{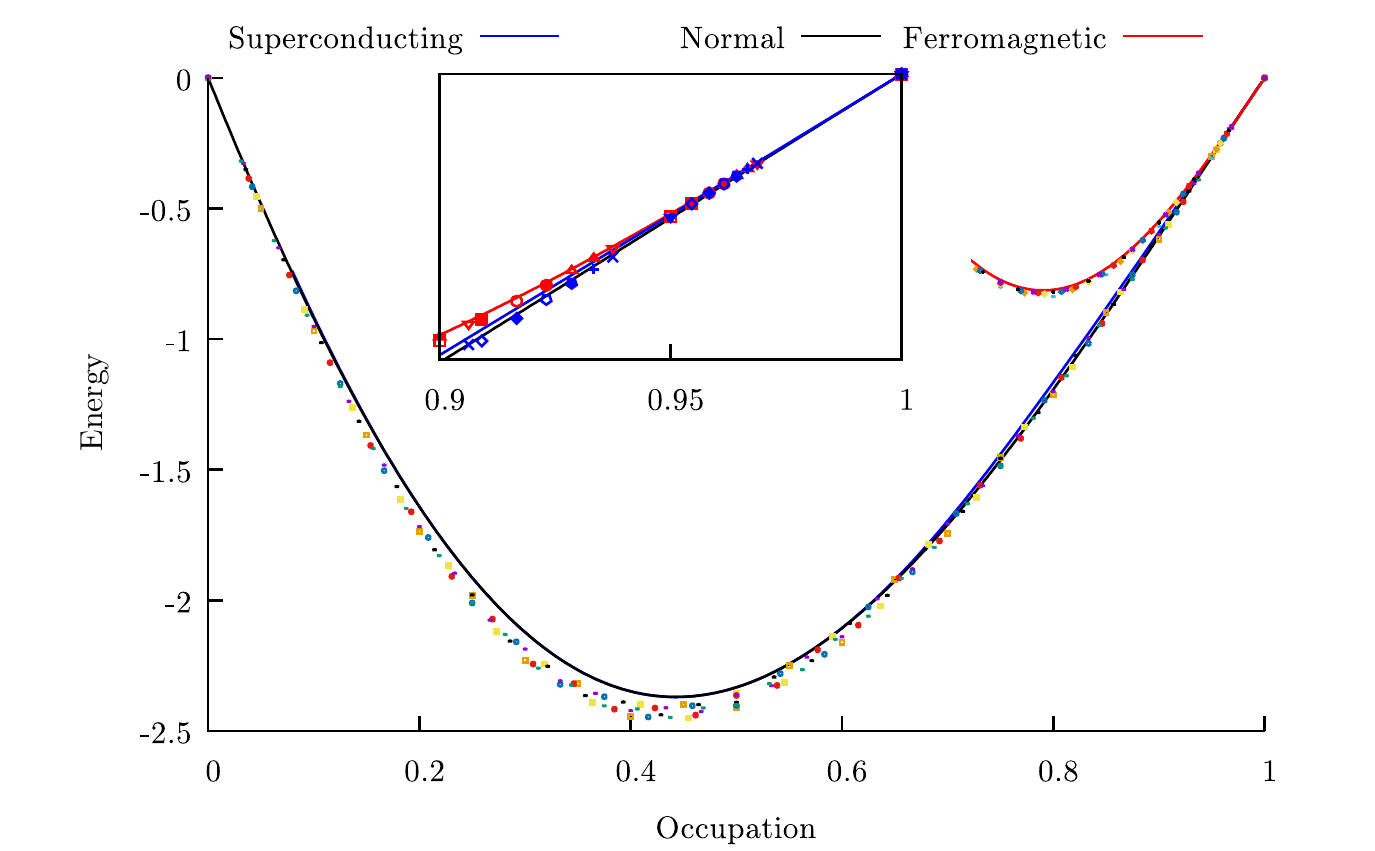}
      \caption{Energy per site per spin as a function of occupation, comparing mean field theory to diagonalisation of finite systems of size 10 to 14 with \(t_0 = t_1\) with \(t_1 = 1\). 
      Superconductivity, paramagnetism, and ferromagnetism are competitive close to the Mott point, however superconductivity is the ground state of the system.
      As there is no competition between the subspaces, this is a pure example of a superconducting system.}\label{fig:energy3}
    \end{figure}

	Figures~\ref{fig:maxwell-construction},~\ref{fig:energy1},~and \ref{fig:energy3} depict the average energy per spin per site for a variety of systems. 
	Mean field results are compared against exact diagonalization of finite sized systems. 
	The anti-symmetric subspace diagonalization results agree extremely well with the mean field results, and when examining the total spin of the system we find it to be a ferromagnet for every choice of parameters.
	The symmetric subspace is slightly more complicated.
	At low occupation every system is a normal metal, but at some point each system begins to favour the superconducting solution. 
	This is complemented by the numerical results which have better agreement with the superconducting solution than the paramagnetic one.
   
    \subsection{Pair Formation}
\begin{figure}
      \centering
      \includegraphics[width = \linewidth]{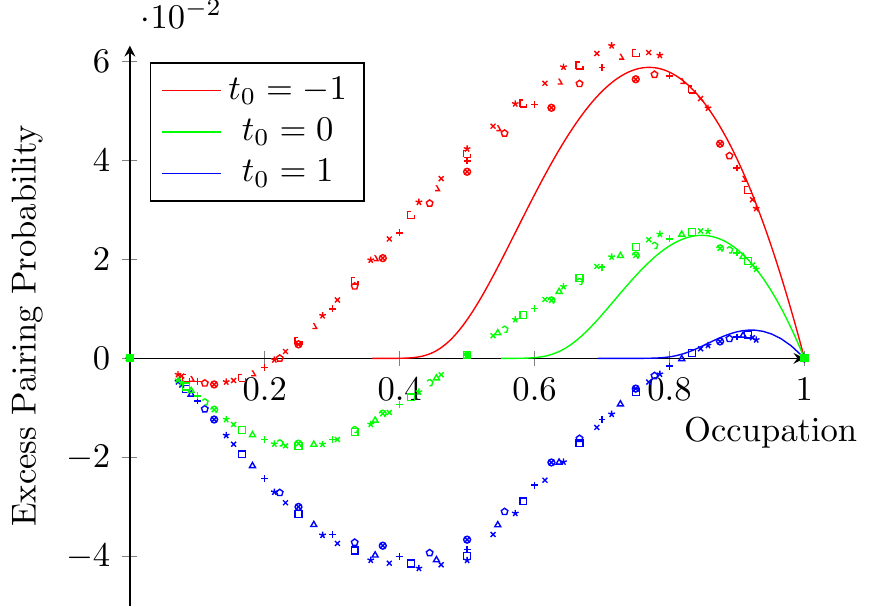}
      \caption{Excess pairing probability, P, as a function of occupation, comparing mean field theory to diagonalisation of finite systems of size 10, 11, 12, 13 and 14 for various values of \(t_0\) with \(t_1 = 1\).
      Close to the Mott point the superconducting mean field theory agrees with diagonalisation results.
      Where the system is repulsively correlated, at low occupation, mean field theory fails to provide an accurate description --- as usual.}\label{fig:pairs}
    \end{figure}

    In conventional superconductivity pairs of electrons form, proliferate, and condense; the number of pairs of electrons in a superconductor is higher than a standard metal.
    We therefore measure the pair formation over an uncorrelated system given by 
    \begin{equation}\label{eq:pairingeq}    
      P = \frac{1}{N} \sum_{i} \Big\langle \big(c^{\dagger}_{i\uparrow} c_{i\uparrow} - \langle c^{\dagger}_{i\uparrow} c_{i\uparrow} \rangle\big)\big(c^{\dagger}_{i\downarrow} c_{i\downarrow} - \langle c^{\dagger}_{i\downarrow} c_{i\downarrow} \rangle\big)\Big\rangle,
    \end{equation}  
    where \(P\) counts the number of pairs in excess of uncorrelated.
    This is depicted in figure~\ref{fig:pairs} and shows that excess pairing, and therefore superconductivity, strengthens with the reduction of \(t_0\). 
    For low occupation the system is strongly correlated against pair formation as discussed in figure~\ref{fig:correlatedmotion}.
    Mean field theory's failure for these types of systems is well known, and so the best it can do is be zero in this region.
    For higher occupation the system prefers pair formation, which agrees with the BCS superconducting solution.
    This is a local quantity and as a result finite size scales well.
    
    \subsection{Superconducting Gap}
\begin{figure}
    	\centering
    	\includegraphics[width = \linewidth]{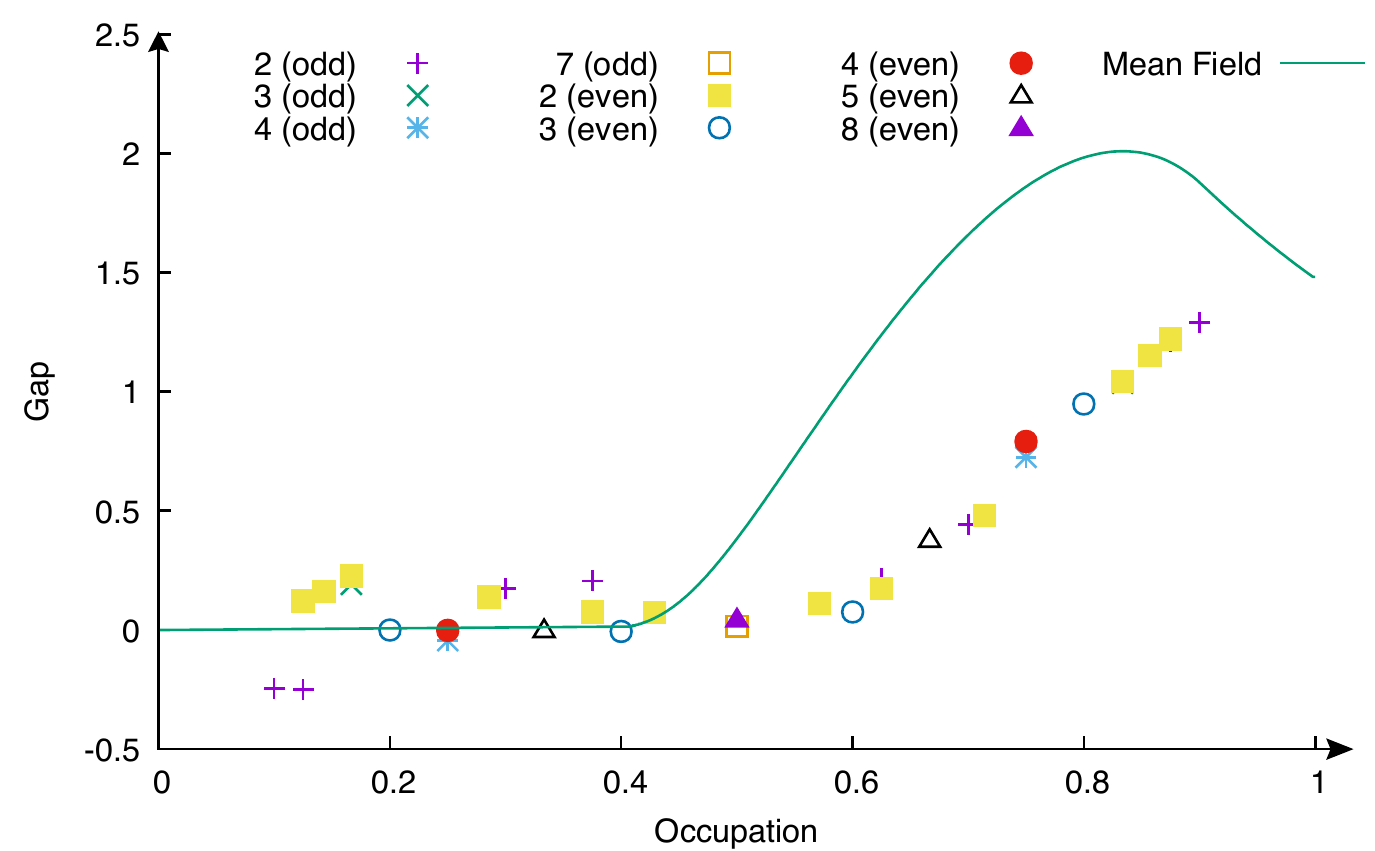}
    	\includegraphics[width = \linewidth]{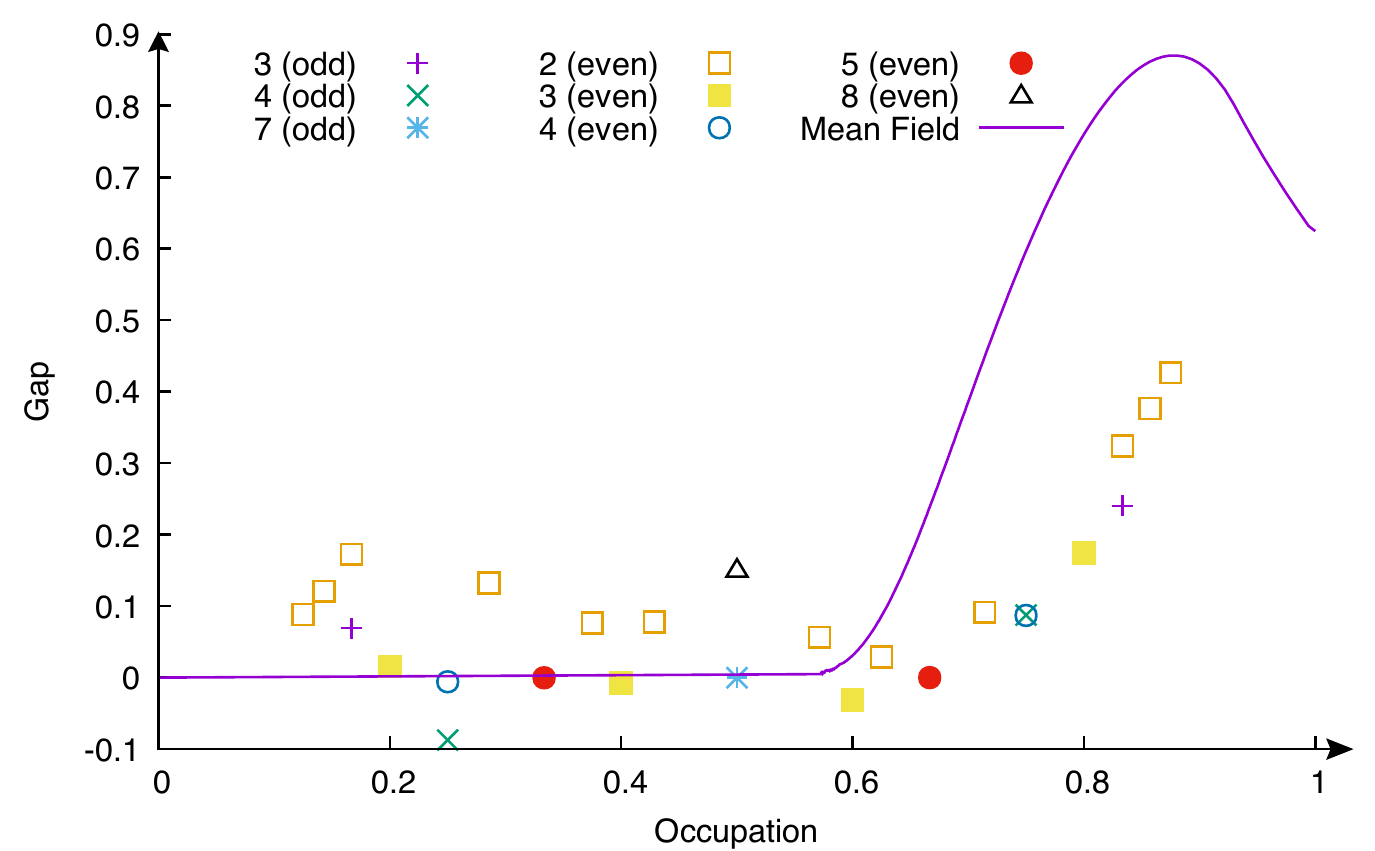}
    	\includegraphics[width = \linewidth]{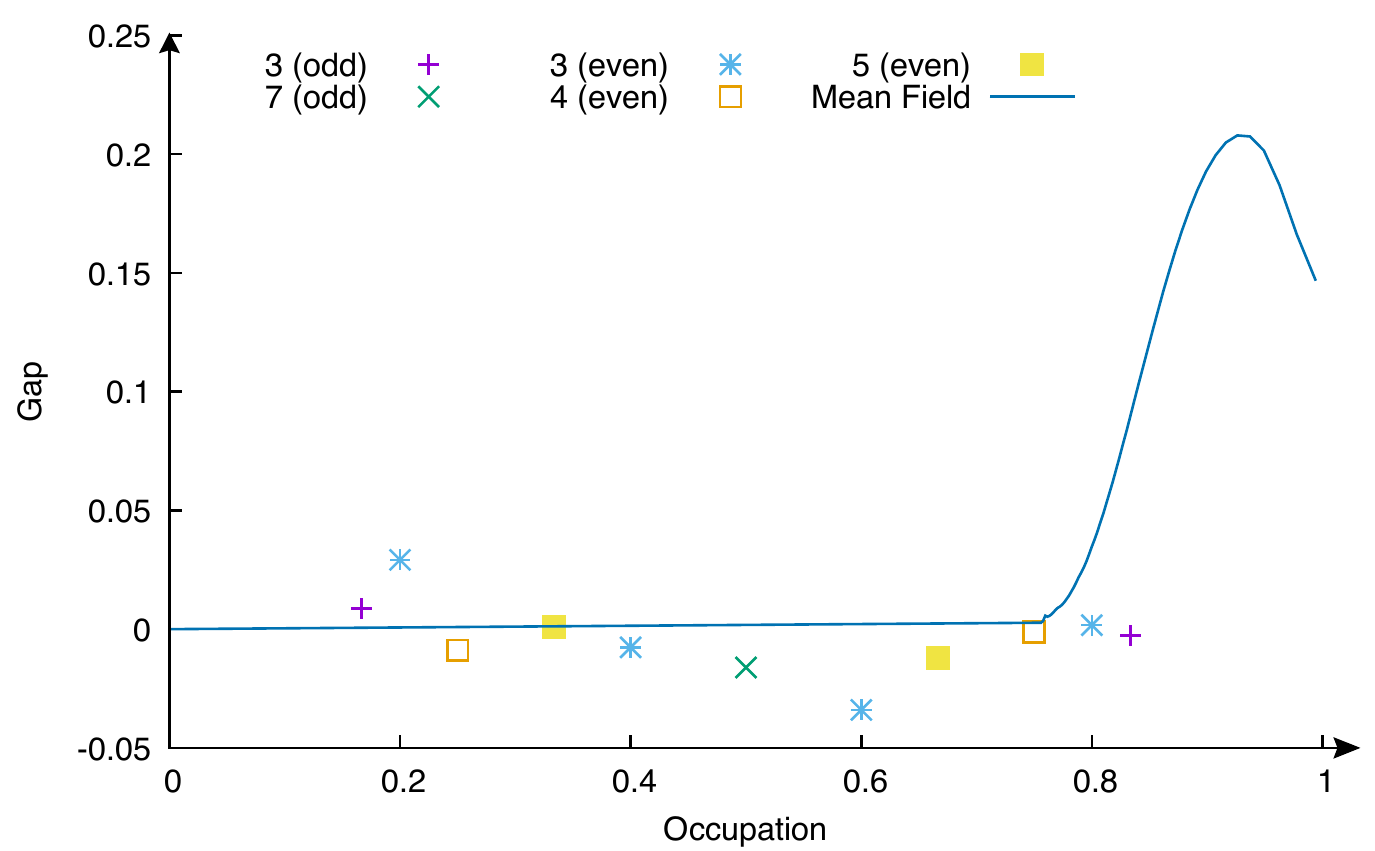}
    	\includegraphics[width = \linewidth]{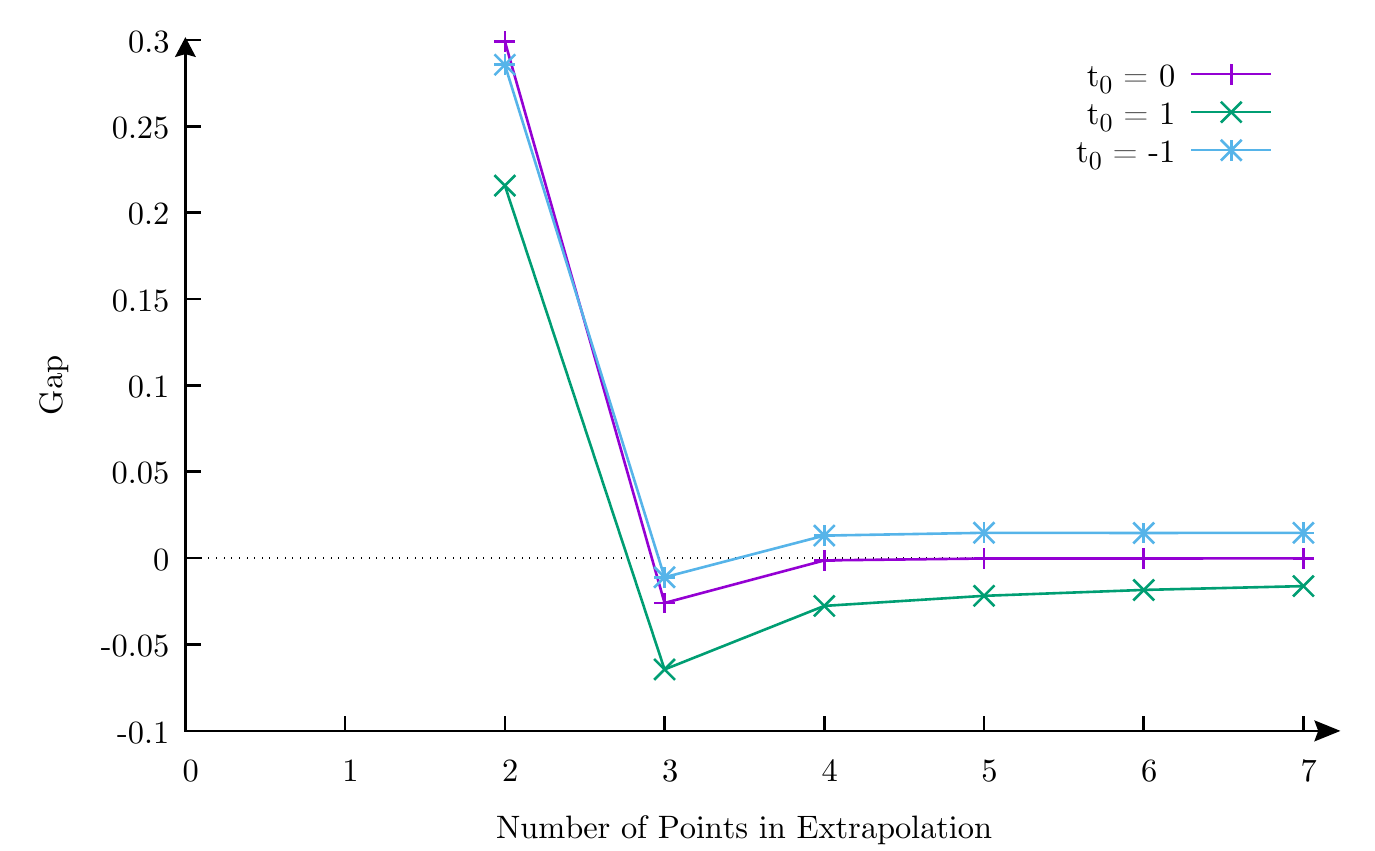}
      \caption{Superconducting gap as a function of occupancy, comparing mean field theory to polynomial extrapolation of diagonalisation of finite systems, with various \(t_0\).
      The number of points we can extrapolate from is a function of occupation as certain ratios only occur for particular system sizes. The first three figures are the gaps for systems with \(t_0 = -1\), 0, and 1. The convergence of extrapolation is depicted in the final figure.}\label{fig:gap}
    \end{figure}

    A cornerstone of superconductivity is the superconducting gap: the excess energy gained from pair formation.
    We calculate the gap from exact diagonalisation by comparing the energy difference for even and odd particles
    \begin{equation}
    	\Delta_N = |E_{N-1} - 2E_{N} + E_{N+1}|.
    \end{equation} 
    This is depicted in figure~\ref{fig:gap}.
    This is the most sensitive calculation of all in this paper as the gap is global property of the system.
    It is incredibly sensitive to occupation and system size, and as a result we use finite size extrapolation to infer how an infinite system would behave.    
    As each occupation ratio may only be attained with certain system sizes we are extrapolate with differing, but the maximal, number of points for each occupation.
    The gap has good agreement with the mean field solution.
    Finally, the Mott point agrees incredibly well and tends to the bound hole-pair state energy, showing the strength of the calculation at this point.
    Experimentally the superconducting transition temperature is directly related to the size of the gap.
    For systems which superconduct we find the transition temperature would be of order 100K, similar to those seen in experiments.
    
    The mean field results show the symmetric subspace is paramagnetic at low occupation and superconducting close to the Mott point, while the anti-symmetric subspace is ferromagnetic for all occupation; all while showing good agreement to diagonalisation results.
    We also found \(t_0\) controls the strength of superconductivity and ferromagnetism.
    If the symmetric and anti-symmetric systems compete energetically, the true ground state is a phase separated mixture of the two subspaces, which is accurately described by a Maxwell construction.
    For \(t_0 = 1\) the symmetric subspace is the true ground-state for all occupation, hence is an example of a system that, without question, displays superconductivity.

 \section{Physical Extensions}\label{sec:physicalExtensions}
    In this section we extend our analysis to more physical systems, namely increasing dimensionality and reducing divergent Coulomb repulsion, and show that superconductivity persists.
    The calculations preceding this section were carried out in one dimension for practical reasons: the numerical results finite size scale more effectively.
    However, in section~\ref{sec:2d}, we will extend our work onto the two dimensional square lattice and find no qualitative differences.
    Separately, the divergent limit of \(U \rightarrow \infty\) was taken in order to remove an energy scale, but in physical systems coulomb repulsion is usually only an order of magnitude higher than chemical bonding.
    Therefore, in section~\ref{sec:UnotInf}, we shall lift this limit perturbatively and find the only difference to be the emergence of an anti-ferromagnetic phase close to the Mott point.
    
      \subsection{2D Square Lattice}\label{sec:2d}
\begin{figure}
          \centering
          \includegraphics[width = \linewidth]{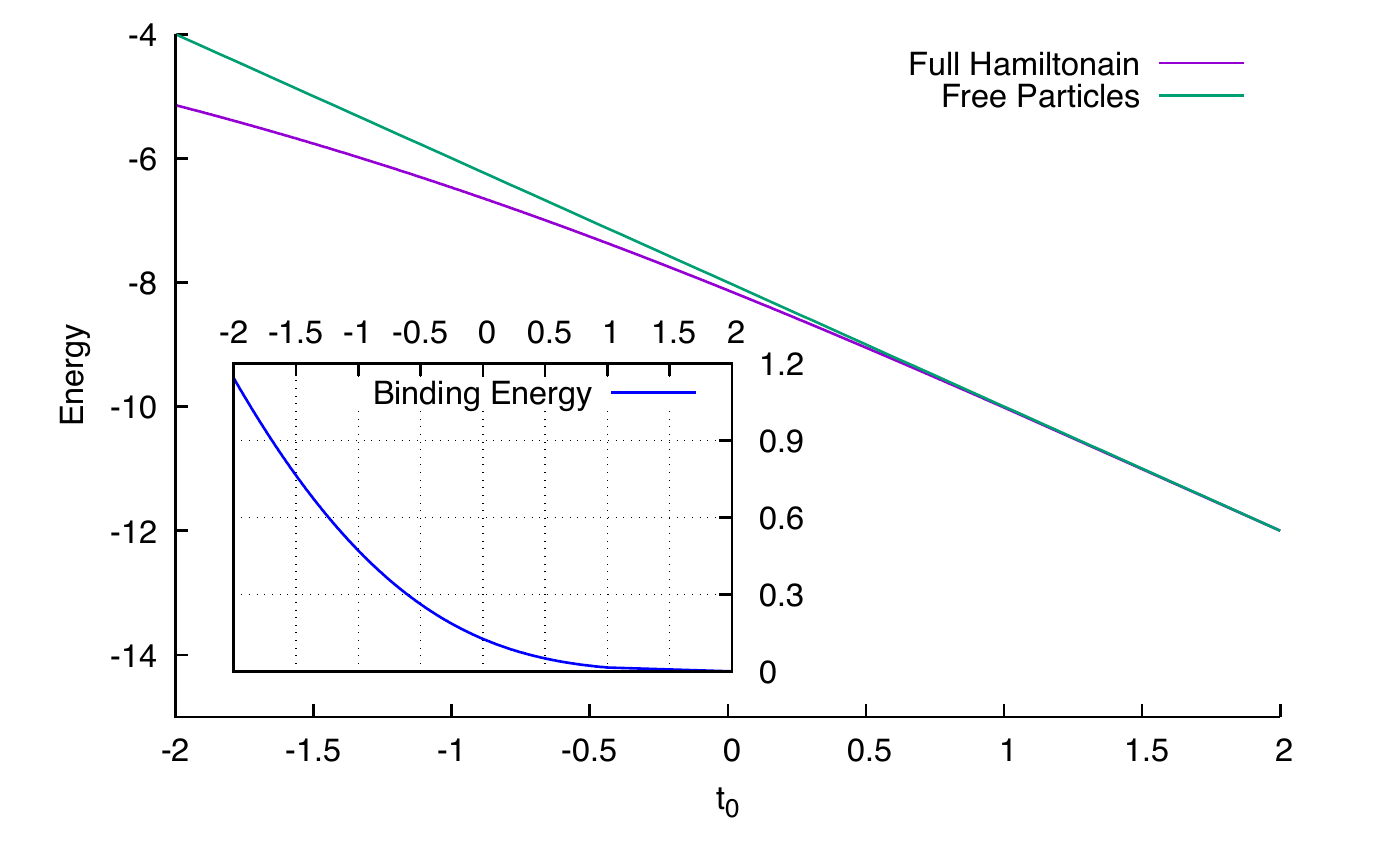}
          \caption{Total energy for two particles with Hamiltonian~\ref{eq:impurity-h0}, labelled free particles, and exact solution labelled full Hamiltonian. The difference of the two is the binding energy. Calculations performed on the 2D square lattice. The binding energy at \(t_0 = 2\) is \(\approx 2 \times 10^{-2}\).}\label{fig:2dimpurity}
        \end{figure}

        The 2D square lattice is the natural structure to extend to, as it is the structure of interest in cuprate superconductors.
        By extending in this way we also are in a regime permitting long range order via the Mermin-Wagner theorem~\cite{MerminWagner}.
        The numerics are calculated using helical boundary conditions~\cite{helical}.
        Unfortunately this severely limits the number of systems we can examine, and finite sized scaling suffers as a result.
        Fortunately, quantities such as the on-site excess pairing and total energy show good agreement in spite of this limitation.
        Unfortunately, the superconducting gap cannot be finite size scaled for, and we only have one point to compare against: the exact solution.
        The change in the analysis is the structure factor becomes that of the 2D square lattice \( \gamma_k = \frac{1}{2}(\cos(k_x) + \cos(k_y))\).
        We can trivially repeat the pairing calculation, but we are now required to perform an elliptic integral.
        The binding energy equation is given by
        \begin{align}    
          \frac{1}{N} \sum_{\mathbf{k}}\frac{1}{\epsilon - 16t_1\gamma_{\mathbf{k}} - 2t_0} = \frac{2}{\epsilon + 4 t_0}.
        \end{align} 
        An analytical expression for \(\epsilon\) cannot be found due to the elliptic integral, therefore it must be solved for numerically.
        This is depicted in figure~\ref{fig:2dimpurity}.
        Qualitatively there is no difference between the 2D and 1D answer.
        Two holes in a fully occupied background still form a bound pair for a large range of \(t_0\).                                   
        
        The mean field calculation is performed in the same manner as before and we find the superconducting energy per spin per site to be
        \begin{multline}
          \bar{E}^{2D}_{SC} =- 16 t_1 n_1 ((1-\eta n_0)^2 - \eta^2 n_1^2) - 2 t_0 n_0 (1 - n_0)  \\
            + 2t_0 \delta_0^2 + 16 t_1 \eta (2\delta_0\delta_1(1-\eta n_0) + \eta n_1(\delta_0^2 + \delta_1^2)),
        \end{multline}    
        while for paramagnetism we have \(\bar{E}^{2D}_{P} = \bar{E}^{2D}_{SC}|_{\delta = 0}\).
        
\begin{figure}
          \centering
          \includegraphics[width = \linewidth]{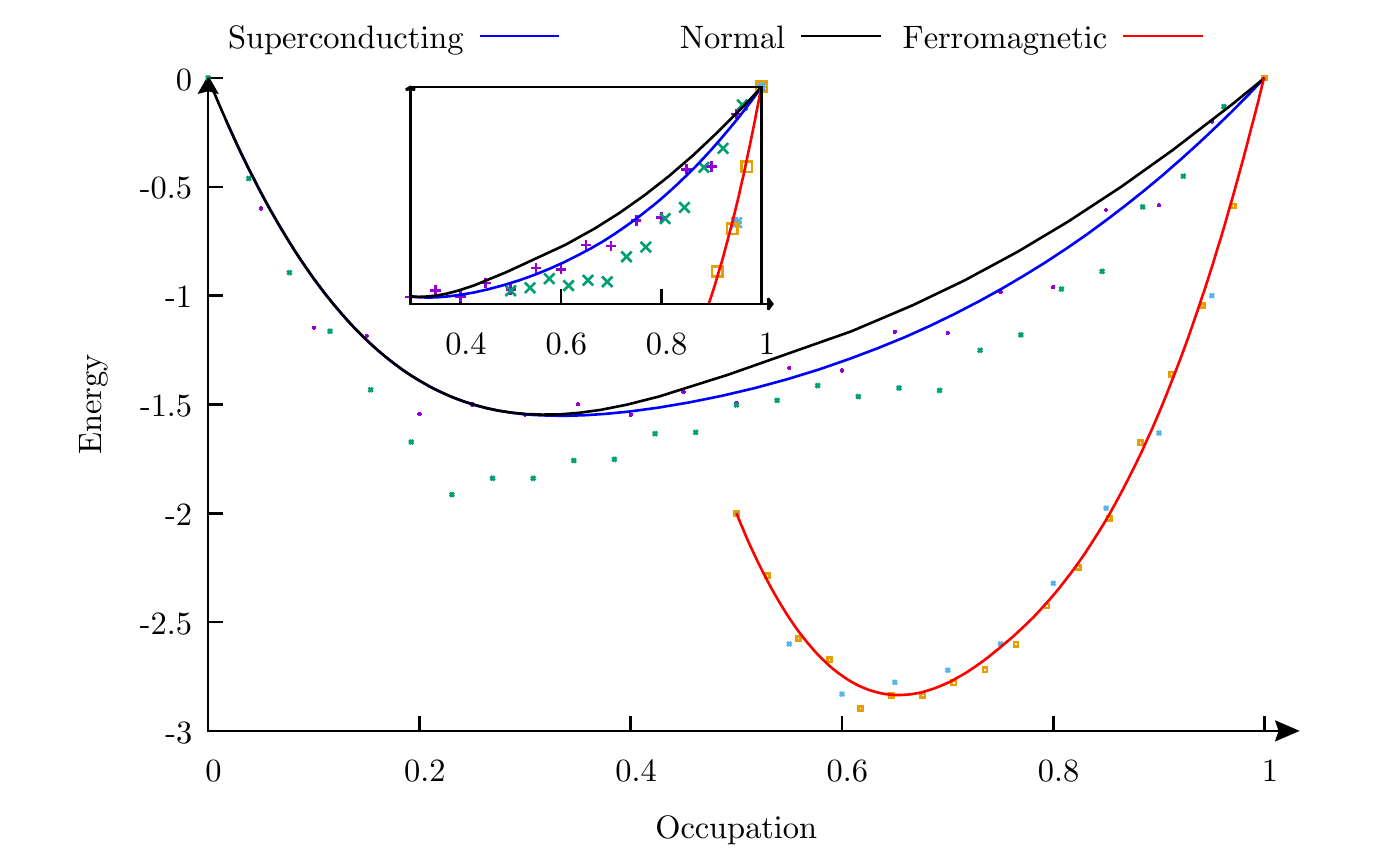}
          \caption{Energy per spin per site as a function of occupation, comparing mean field theory to diagonalisation of a finite system of size \(3\times 3\) and \(3 \times 4\) with \(t_0 = -2\) and \(t_1 = 1\).
          Superconductivity has better agreement over paramagnetism, both existing within the symmetric subspace.
          The anti-symmetric subspace is ferromagnetic and energetically dominant in the region where superconductivity is prevalent, hence this system is not superconducting.}\label{fig:2d-2}
        \end{figure}

\begin{figure}
          \centering
          \includegraphics[width = \linewidth]{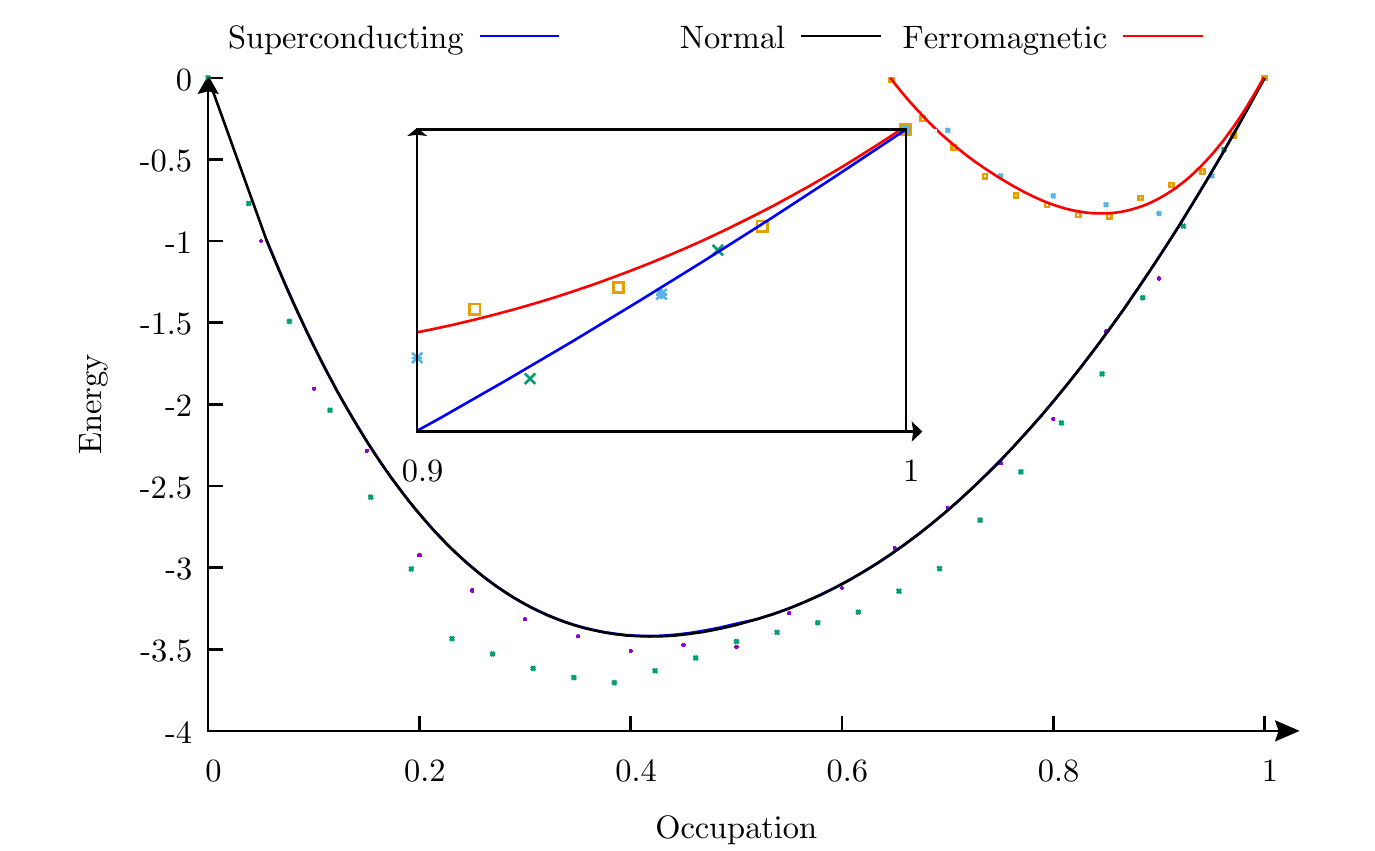}
          \caption{Energy per spin per site as a function of occupation, comparing mean field theory to diagonalisation of a finite system of size \(3\times 3\) and \(3 \times 4\) with \(t_0 = 2\) and \(t_1 = 1\).
          Superconductivity is the favourable phase close to the Mott point where it competes with paramagnetism and ferromagnetism.
          This is an example of a 2D system which is superconducting.}\label{fig:2d2}
        \end{figure}

        Again our intuition about the anti-symmetric subspace is correct and we find it to be ferromagnetic from numerical calculations.
        The difference in energy between the superconducting state and the paramagnetic state is smaller than before and is depicted in figure~\ref{fig:2d-2}.
        Just as before the competition between subspaces means only particular systems are  described by the symmetric subspace for all occupation; figure~\ref{fig:2d2} is an example of one.   
        
\begin{figure}
          \centering
          \includegraphics[width = \linewidth]{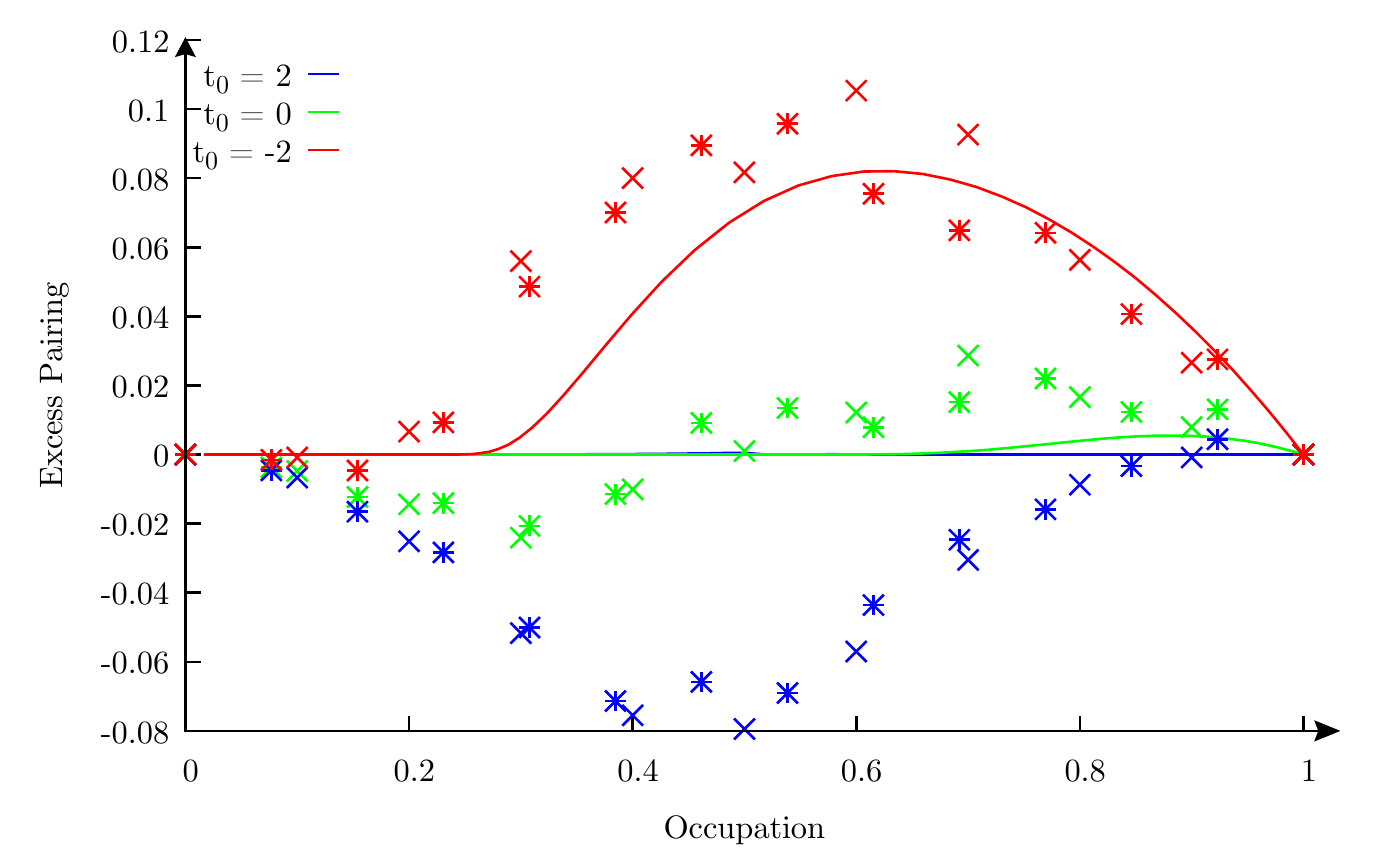}
          \caption{Excess pairing probability, P, against occupation for the 2D square lattice with \(t_1 = 1\) and varying \(t_0\), comparing mean field theory to diagonalisation of a finite system of size \(3\times 3\) and \(3 \times 4\).
          Mean field theory shows good agreement to diagonalisation results, where superconductivity exists close to the Mott point.}\label{fig:2dpairs}
        \end{figure}

        Considering pairing at other occupation we see, using mean field theory and numerics, superconducting hole pair formation close to the Mott point, this is depicted in figure~\ref{fig:2dpairs}.     
        
	    \begin{figure}
	    	\centering
	    	\includegraphics[width = \linewidth]{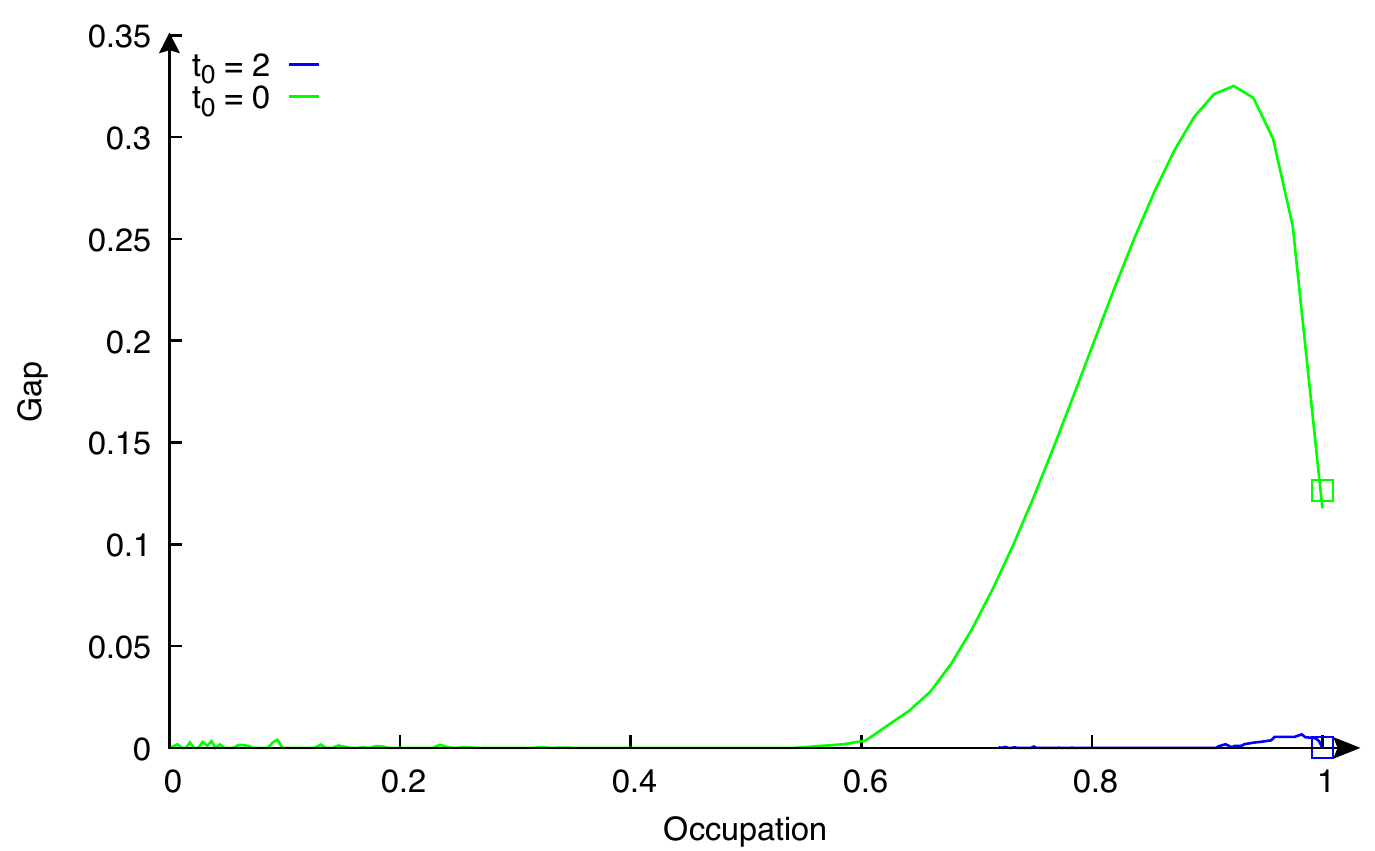}
	    	\caption{Superconducting gap, against occupation for the 2D square lattice with \(t_1 = 1\) and varying \(t_0\). We are unable to compare against finite sized diagonalisation as there is not enough data to finite size scale. However, the exact solution at the Mott point agrees well.}\label{fig:2dgap}
	    \end{figure}

      \subsection{Lifting the Limit \(U \rightarrow \infty\)}\label{sec:UnotInf}
        A key requirement in the previous analysis was the divergent limit of \(U\) as it gave access to exact results. 
        In order to examine whether or not the superconductivity is unique to \(U \rightarrow \infty\), we lift the limit in this section.
        The numerical calculations from section~\ref{sec:results} are trivially repeated; however, the increased local state-space reduces the maximum system size.     
        Analytically we are restricted to the limit \(U\) is large but not infinite, where we can perturbatively expand the Hamiltonian in terms of order \(t^2/U\), more commonly known as the t-J model. 
        Physically this perturbation corresponds to virtual hopping on to states that would cost \(U\) in order to gain from hopping energy.
        In the anti-symmetric subspace things are more difficult, as allowing perturbative hopping affects the spin state.
        In section~\ref{sec:nonlinear} an argument from energy was made to describe the system as ferromagnetic, however the virtual states now accessible cause a reduction in the spin.
        This is a manifestation of the Haldane gap~\cite{Haldane}, where the ground state of the quantum spin-half ladder at the Mott point is given by a valence bond on each site.
        Numerically we find the system exhibits a zero-temperature phase transition where the spin changes from maximal to zero.
        Due to the intricate nature of the anti-symmetric subspace we are unable to provide any approximate analysis for it but find it to be described by an anti-ferromagnet close to the Mott point.
        
        Performing the perturbative expansion on the symmetric subspace Hamiltonian provides the average energy as given in equation~\ref{eq:U-ham}, where \(n_2 = \langle c^\dagger_{i,\sigma}c_{i\pm 2,\sigma}\rangle \), the next nearest neighbour occupation is now introduced.
        The paramagnetic average energy per spin per site is given by \(\bar{E}_{P_U} = \bar{E}_{SC_U}|_{\delta = 0} \). 
        All relevant parameters can be found in the same way as before, by obtaining self consistent integral equations.
        
\begin{figure}
          \centering
          \includegraphics[width = \linewidth]{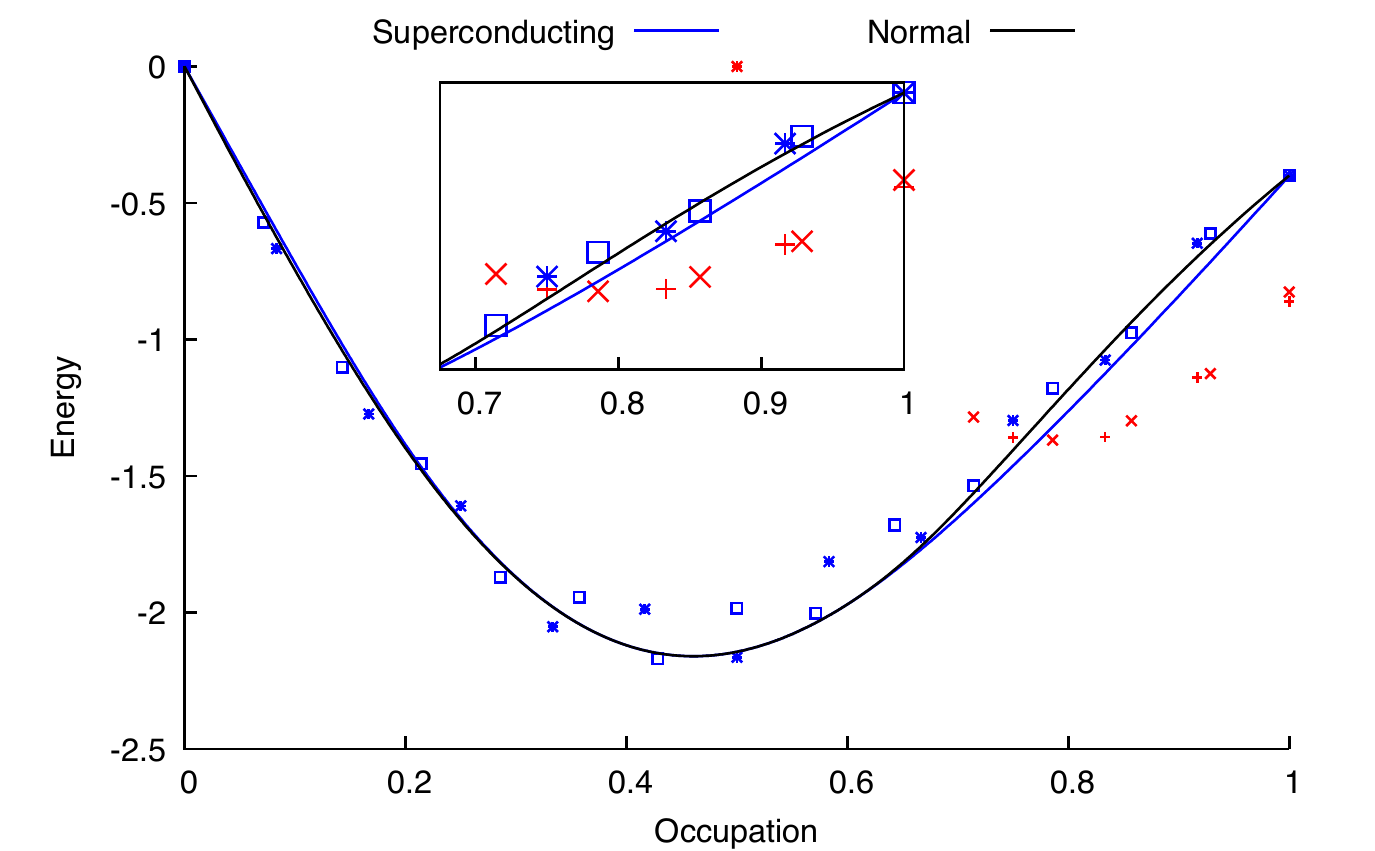}
          \caption{Energy per spin per site as a function of occupation, comparing perturbative mean field theory to exact diagonalisation of finite systems of size 8 and 9 with \(U = 10\), \(t_0 = 0\) and \(t_1 = 1\).
          The anti-symmetric subspace cannot be characterized, however there is a phase transition from ferromagnetism to anti-ferromagnetism close to the Mott point.
          Mean field theory has good agreement to the diagonalisation data, with superconductivity being favourable over paramagnetism but not the anti-symmetric subspace.}\label{fig:U1}      
        \end{figure}

        Figure~\ref{fig:U1} demonstrates a case where the symmetric subspace is the true ground state.
        Again in this system superconductivity occurs close to the Mott point.
        In these results the perturbative expansion consistently produces an overestimate for the energy gained due to virtual hopping.
        In principle this can be remedied by adding higher order terms such as \(t^3/U^2\), but it is not done in this paper.
        
\begin{figure}
          \centering
          \includegraphics[width = \linewidth]{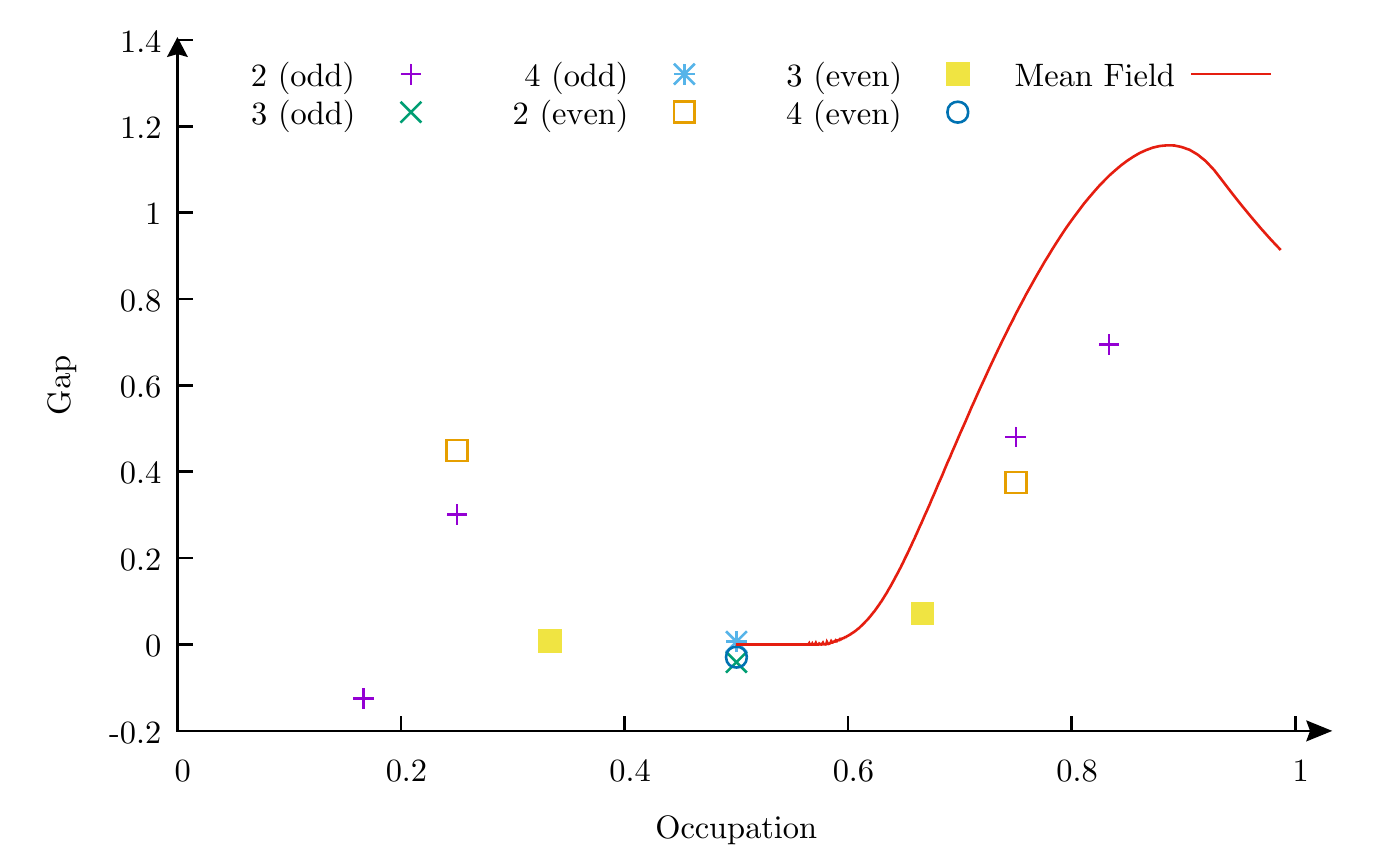}
          \includegraphics[width = \linewidth]{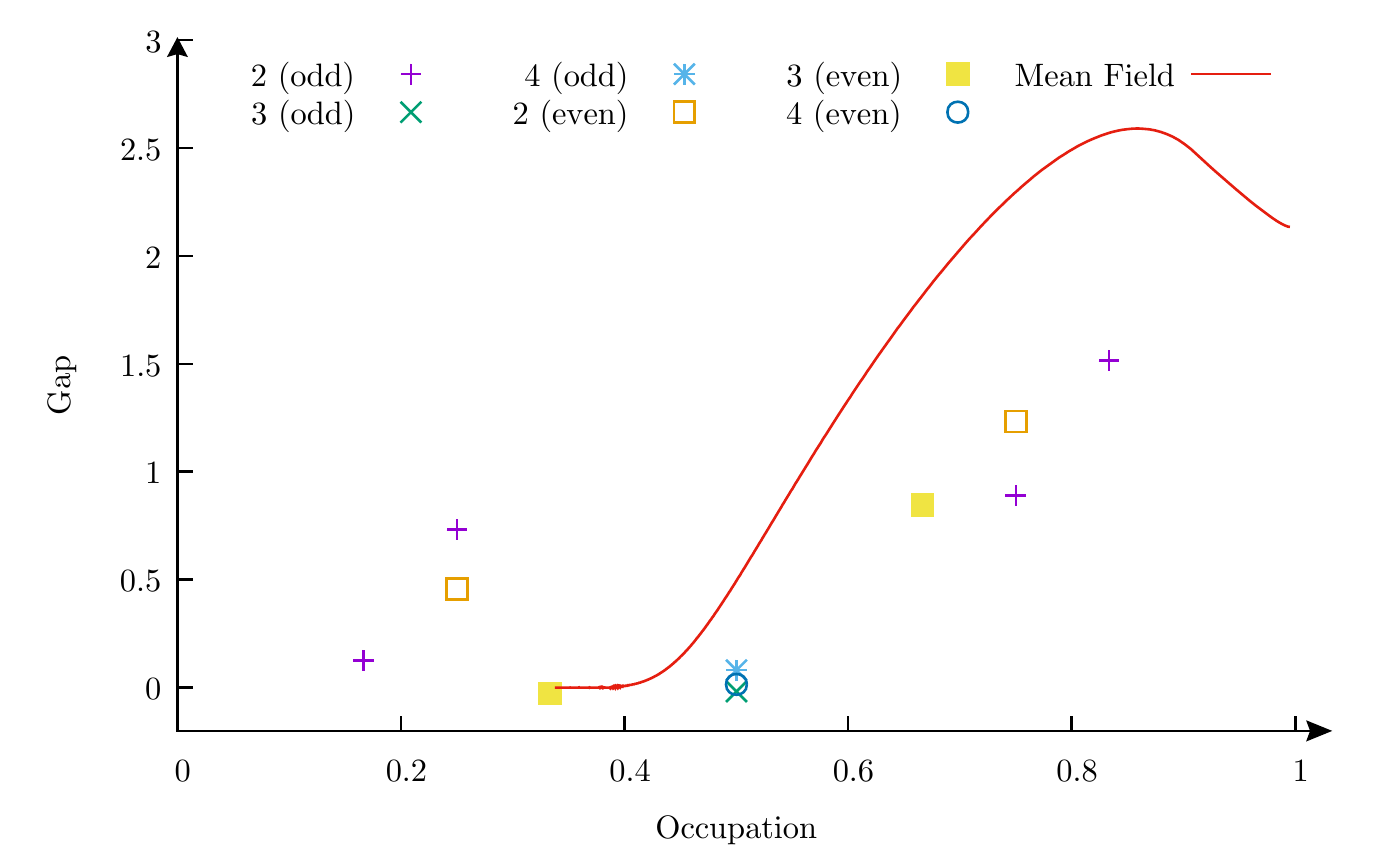}
          \caption{Superconducting gap of a system with \(U = 10\), and \(t_0 = 0\) and \(-1\). We only have access to smaller systems and hence have limited finite sized scaling. Despite this there is still good agreement.}\label{fig:UPairs} 
        \end{figure}

        The superconducting gap is calculated in the same way as previous calculations and is depicted in figure~\ref{fig:UPairs} .
        Again we are limited by system size for extrapolation, but the trend seems to agree well with the mean field data.
        \begin{widetext}
        \begin{multline} \label{eq:U-ham}
          \bar{E}_{SC_{U}} = -\frac{8 {t_1}^2}{U} \Bigg[\frac{1}{2} \Big[{n_0}(-({\delta_0}^2+{\delta_1}^2))
          +2  {\delta_0} {\delta_1} {n_1}-{n_0} ({n_0}^2-{n_1}^2)\Big]+
            \Big[{\delta_1}^2 +{n_1}^2\Big] \Big[\eta ^2({\delta_0}-{\delta_2})^2+(1-\eta ({n_0}-{n_2}))^2\Big] \\
              +\eta ^2 \Big[{\delta_0}-{\delta_2}\Big]^2 \Big[1-\eta  ({n_0}-{n_2})\Big]\Big[{n_1}^2-{n_0} {n_2}\Big] \Big[2{\delta_1} \eta  {n_1} ({\delta_0}+{\delta_2}) 
             -2 {\delta_0}{\delta_2} \eta  {n_0}+({\delta_1}^2+{n_0}{n_2}) (1-\eta  ({n_0}-{n_2}))+2 \eta {n_1}^2 {n_2}\Big]\\
             +2{\delta_1}^2+{n_0}^2+{n_1}^2\Bigg] 
             -8 {t_1} \Bigg[{n_1} \Big[(1-\eta  {n_0})^2-\eta ^2({\delta_0}^2+{\delta_1}^2+{n_1}^2)\Big]-2{\delta_0} {\delta_1} \eta  \Big[1-\eta  {n_0}\Big]\Bigg] \\
             -\frac{8\sqrt{2} {t_0} {t_1} }{U}\Bigg[-\eta  \Big[-{n_1}({\delta_0}^2+{\delta_1}^2)+2 {\delta_0} {\delta_1}{n_0}+{n_1}({n_0}^2-{n_1}^2)\Big]
             +{\delta_0}{\delta_1}+{n_0} {n_1}\Bigg] 
             -\frac{4 {t_0}^2}{U}\Bigg[{\delta_0}^2+{n_0}^2\Bigg] -2 {t_0} \Bigg[(1-{n_0}) {n_0}-{\delta_0}^2\Bigg]    
        \end{multline}
        \end{widetext}
        
        \begin{figure}
          \centering
          \includegraphics[width = \linewidth]{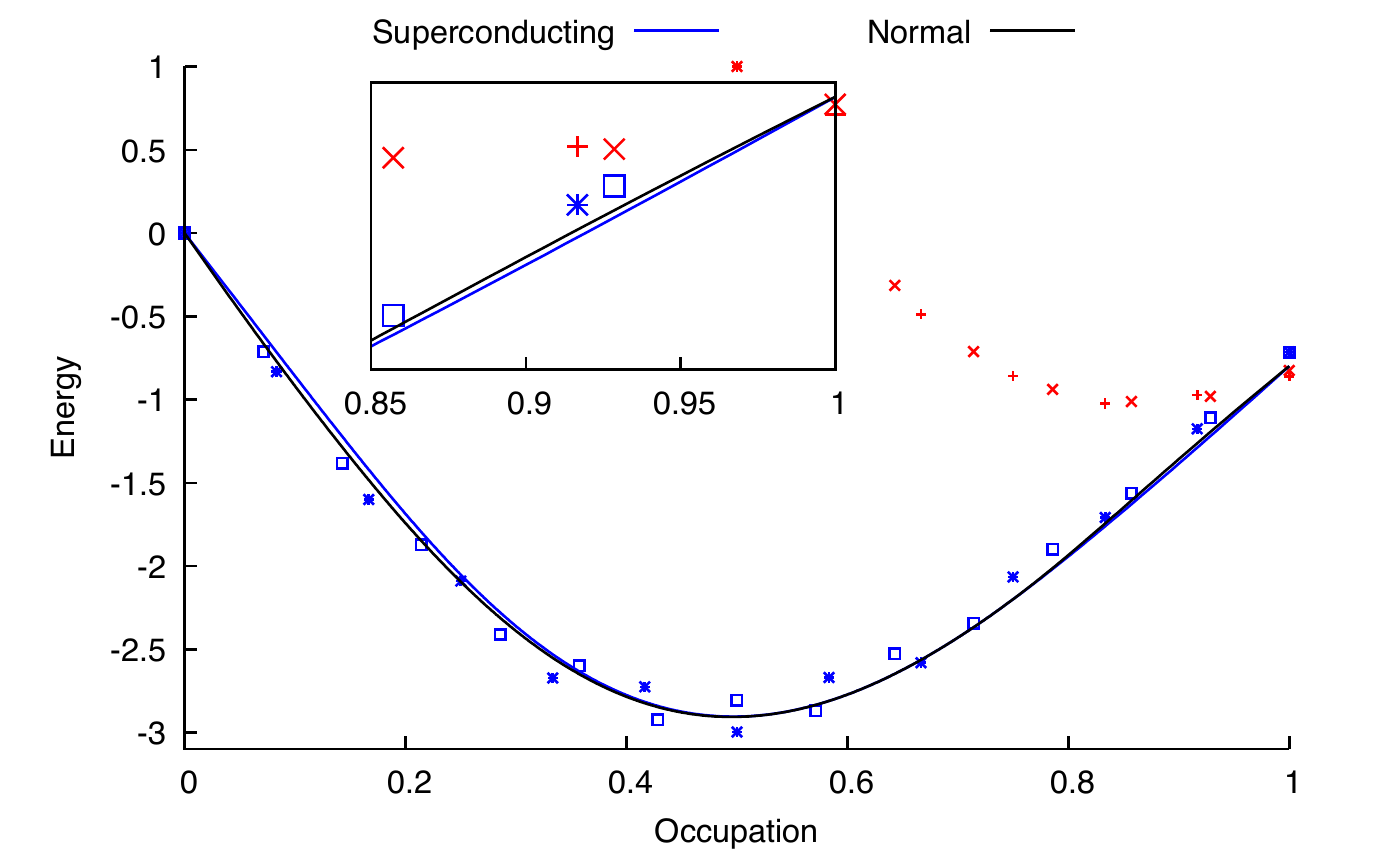}
          \caption{Energy per spin per site as a function of occupation, comparing perturbative mean field theory to exact diagonalisation of a finite system of size 8 and 9 with \(U = 10\), \(t_0 = 1\) and \(t_1 = 1\).
          Superconductivity is the favoured phase, close to the Mott point, while competing with anti-ferromagnetism (from the anti-symmetric subspace) and paramagnetism (from the symmetric subspace).}\label{fig:U2} 
        \end{figure}
        
  \section{Conclusions}
    At the most pragmatic level, we have provided a strongly correlated model which is mathematically more tractable than usual. 
    The model exhibits four natural phases; a paramagnet, a superconductor, a ferromagnet and an anti-ferromagnet. 
    These phases are very common in strongly correlated systems and we believe that the physical cause of these phases in our model may well be similar to that in the experimental systems. 
    We further believe that the basic mathematical technique, non-linear fermion transformations will shed light on the models which should naturally be derived from the experimental systems. 
    It is clear that the paramagnet at low occupancy arises from the same physical source of Coulomb repulsion between charge carriers. 
   	It is further clear that the anti-ferromagnet also stems from the standard source of kinetic exchange suggested for more elementary Mott insulating models. 
   	The ferromagnetism is analogous to that found in manganites, the immobile \(t_{2g}\) electrons correspond to our passive anti-symmetric electrons and the mobile \(e_g\) electrons have improved conductivity if they align all the \(t_{2g}\) spins. 
   	Obviously there are further complications in the manganites, but the Zener exchange is very similar. 
   	Finally, we have the superconducting phase, and here there is no accepted mechanism for superconductivity in strongly correlated systems and so our model provides such a mechanism.
    
    At very large U, close to the Mott insulator, we find a competition between the ferromagnet and the superconductor with the ferromagnet ultimately winning. 
    This can be overturned using \(t_0\), but this inclusion also weakens the superconductivity.
    Physically, the finite Hubbard repulsion is the likely source of any destabilisation of the ferromagnet and is likely to be crucial in the experimental systems. 
    This effect also stabilises the anti-ferromagnet very close to the Mott insulator and this is of course experimentally observed.
    
    We have rigorously demonstrated the superconductivity within this model. 
    The mean-field theory fits well to the local correlations and predicts the total energy nicely. We are at the same level as the original BCS theory. 
    We can prove that the pairs form in the model close to Mott insulator; this is an exact result. 
    We can even find the dispersion and coherence of the pairs exactly. 
    Firstly, we analysed the one-dimensional variant in detail and we know that low dimensional fluctuations will destroy the phase coherence at long range; a careful investigation would be dogged by the expected weak power-law behaviour. 
    Secondly, we extended this analysis to two dimensions, where phase coherence at long range is permitted.
    There is no reason to believe that our model is abnormal.
    
    Our model exhibits a new mechanism for superconductivity; correlated hopping. 
    The chemical bonding energy depends strongly on the local electron concentration. 
    If the system is locally empty there is a huge potential for bonding, whereas if the system is locally full the potential bonding is severely restricted by the correlations; the avoidance of the Coulomb penalties blocks the majority of the hopping. 
    In our model this competition is non-linear, it is worth forming local number fluctuations to gain from the empty regions because the local configurations with the average global occupancy are so badly blocked and even denser regions are not a lot worse. 
    One might also worry about the possibility of charge density waves being competitive with the superconductivity, but the fact that it is a bonding phenomenon and not an on-site attraction eliminates this common issue.
    
    We have constructed a strongly-correlated model which is mathematically more tractable than usual. 
    The model exhibits a wealth of phases, that are observed in strongly correlated systems, of which the most interesting is superconducting. 
    The pairing is a weak correlation on a basically free-electron foundation and is very credible.
    We now need to look at the scope of the mathematics and whether it can be developed into a standard tool to investigate general strongly-correlated tight-binding models and in particular those that naturally crop up in the study of experimental systems. 
    The underlying technique is that of non-linear fermion transformations. 
    At the simplest level this technique allows us to construct a new fermion from the original set of fermions, but allows a free choice of which states to pick at the single-particle and two-particle levels; we can choose to use any two-particle state independent from the choice of one-particle states. 
    The strength of this option is that we can avoid the Coulomb prohibited states which would naturally occur if we simply doubly occupied the one-particle states. The choice of the best one-particle and two-particle states then becomes an active ingredient in the theory. 
    For our model the symmetry and Coulomb restrictions force this issue and we have no freedom, the effective model is unique. 
    If we want to apply a non-linear fermion transformation to a more general model then we need to make the transformation variable and employ the energetics to decide the best choice of local states, a much more involved procedure which is currently being worked on.
  
  \bibliographystyle{unsrt}
  \bibliography{bibliography}
  
\end{document}